\documentclass{aa}  
\usepackage{threeparttable}
\usepackage{natbib}
\usepackage{xcolor}
\usepackage{placeins}

\usepackage{microtype}
\usepackage{hyperref}
\usepackage{graphicx}
\usepackage{array}
\usepackage{txfonts}

\begin{document}

   \title{The Fornax 3D project: Globular clusters tracing kinematics and metallicities\thanks{The globular cluster catalogue (full Table A.1) is only available
at the CDS via anonymous ftp to \url{cdsarc.u-strasbg.fr}
(130.79.128.5) or via \url{http://cdsarc.u-strasbg.fr/viz-bin/cat/J/A+A/637/A26}
}, \thanks{Based on observations collected at the ESO Paranal La Silla Observatory,
Chile, Prog. 296.B-5054(A)}}

\titlerunning{Fornax3D: Globular clusters tracing kinematics and metallicities}

   \author{K. Fahrion\inst{1}
          \and
          M. Lyubenova\inst{1}
          \and
          M. Hilker\inst{1}
          \and
           G. van de Ven\inst{2} 
           \and
          J. Falc\'{o}n-Barroso\inst{3}\fnmsep\inst{4}          
          \and
          R. Leaman\inst{5}
          \and
           I. Mart\'{i}n-Navarro\inst{6}\fnmsep\inst{4}\fnmsep\inst{7}\fnmsep\inst{5}
          \and
          A. Bittner\inst{1}
          \and
           L. Coccato\inst{1}
           \and
           E. M. Corsini\inst{8}\fnmsep\inst{9}         
           \and
           D. A. Gadotti\inst{1}
          \and
          E. Iodice\inst{1}\fnmsep\inst{10}
           \and           
          R. M. McDermid\inst{11}
          \and 
          F. Pinna\inst{5}
          \and
           M. Sarzi\inst{12}\fnmsep\inst{13}
           \and
           S. Viaene\inst{14}
          \and
          P. T. de Zeeuw\inst{15}\fnmsep\inst{16}
          \and
          L. Zhu\inst{17}}

   \institute{European Southern Observatory, Karl-Schwarzschild-Stra\ss{}e 2, 85748 Garching bei M\"unchen, Germany\\
   				\email{kfahrion@eso.org}
         \and 
             Department of Astrophysics, University of Vienna, T\"urkenschanzstrasse 17, 1180 Wien, Austria
             \and 
			Instituto de Astrof\'isica de Canarias, Calle Via L\'{a}ctea s/n, 38200 La Laguna, Tenerife, Spain
			\and 
			Depto. Astrof\'isica, Universidad de La Laguna, Calle Astrof\'isico Francisco S\'{a}nchez s/n, 38206 La Laguna, Tenerife, Spain
             \and 
             Max-Planck-Institut f\"ur Astronomie, K\"onigstuhl 17, 69117 Heidelberg, Germany
               \and 
               Instituto de Astrof\'{i}sica de Canarias, E-38200 La Laguna, Tenerife, Spain       
			\and 
               University of California Santa Cruz, 1156 High Street, Santa Cruz, CA 95064, USA
               \and 
                Dipartimento di Fisica e Astronomia 'G. Galilei', Universit\`a di Padova,  vicolo dell'Osservatorio 3, I-35122 Padova, Italy
			\and 
			INAF--Osservatorio Astronomico di Padova, vicolo dell'Osservatorio 5, I-35122 Padova, Italy
			 \and 
             INAF-Astronomical Observatory of Capodimonte, via Moiariello 16, I-80131, Napoli, Italy
               \and 
             Department of Physics and Astronomy, Macquarie University, North Ryde, NSW 2109, Australia
             \and 
             Armagh Observatory and Planetarium, College Hill, Armagh, BT61 9DG, Northern Ireland, UK
             \and 
             Centre for Astrophysics Research, University of Hertfordshire, College Lane, Hatfield AL10 9AB, UK
             \and 
             Sterrenkundig Observatorium, Universiteit Gent, Krijgslaan 281, B-9000, Gent, Belgium
             \and 
             Sterrewacht Leiden, Leiden University, Postbus 9513, 2300 RA Leiden, The Netherlands
             \and 
             Max-Planck-Institut f\"ur extraterrestrische Physik, Gie\ss{}enbachstraße 1, 85748 Garching bei M\"unchen, Germany
             \and 
             Shanghai Astronomical Observatory, Chinese Academy of Sciences, 80 Nandan Road, Shanghai 200030, China
             }

   \date{}

 
  \abstract
  {Globular clusters (GCs) are found ubiquitously in massive galaxies and due to their old ages, they are regarded as fossil records of galaxy evolution. Spectroscopic studies of GC systems are often limited to the outskirts of galaxies, where GCs stand out against the galaxy background and serve as bright tracers of galaxy assembly. In this work, we use the capabilities of the 
Multi Unit Explorer Spectrograph (MUSE) to extract a spectroscopic sample of 722 GCs in the inner regions ($\lesssim 3 R_\text{eff}$) of 32 galaxies in the Fornax cluster. These galaxies were observed as part of the Fornax 3D project, a MUSE survey that targets early and late-type galaxies within the virial radius of Fornax. After accounting for the galaxy background in the GC spectra, we extracted line-of-sight velocities and determined metallicities of a sub-sample of 238 GCs. We found signatures of rotation within GC systems, and comparing the GC kinematics and that of the stellar body shows that the GCs trace the spheroid of the galaxies. While the red GCs prove to closely follow the metallicity profile of the host galaxy, the blue GCs show a large spread of metallicities but they are generally more metal-poor than the host.}
   \maketitle
%

\section{Introduction}
\label{sect:intro}

Globular clusters (GCs) are massive and dense star clusters that are found in almost all types of galaxies, from low-mass dwarfs to the most massive elliptical galaxies (see reviews by \citealt{Brodie2006, Forbes2018b}). With effective radii of a few parsecs and masses between $10^{4}$ and $10^6\,M_\sun$ (e.g. \citealt{Jordan2007, Masters2010}), GCs are bright enough to be observed in distant galaxies, even in the outskirts, where the observation of integrated stellar light is challenging. Typical GCs, such as the ones observed in the Milky Way (MW), have stellar ages \mbox{$\geq$ 10 Gyr} and are among the oldest structures in the Universe (e.g. \citealt{Puzia2005, Strader2006, VandenBerg2013}). These ages suggest a formation redshift $z \gtrsim 2$ and thus their stellar population should reflect the chemical composition of their birthplace. Their orbital properties, in turn, provide hints as to the assembly history and dynamical evolution of their parent galaxy. For this reason, GCs are often used as fossil records of galaxy evolution (e.g. \citealt{Peng2008, Forbes2010, Forbes2011, Brodie2014, Harris2016, Harris2017}).

In the last decades, large photometric surveys, for example with the Advanced Camera for Surveys (ACS) on board the \textit{Hubble Space Telescope} (HST) in dense environments such as the Virgo or Fornax galaxy clusters \mbox{(e.g. \citealt{Cote2004, Jordan2007})}, have collected extensive photometric catalogues of GC candidates \citep{Peng2006, Jordan2015}. Aside from their mass and size (effective radius), GCs are typically classified by their photometric colour (e.g. in the ACS $g$ and $z$-bands) and it has been found that many galaxies have a bimodal GC colour distribution with a red and blue population (e.g. \citealt{Kundu2001, Larsen2001, Peng2006, Sinnott2010}). This colour bimodality has been interpreted as a metallicity bimodality and as indication for a two-phase formation scenario of massive galaxies \citep{Ashman1992, Cote1998, Beasley2002}. Connected to the hierarchical merger scheme of galaxy evolution, metal-rich GCs are assumed to form in-situ in massive haloes, whereas the metal-poor GCs have their origin in gas-rich dwarf galaxies that were later accreted \citep{Kravtsov2005, Tonini2013, Li2014}.

Although photometric studies of GC systems can provide first insights, spectroscopy is needed to effectively use GCs as tracers of galaxy assembly. With spectra, line-of-sight (LOS) velocities of individual GCs can be determined, which are crucial to study the kinematics of the (global) GC system (GCS). Spectroscopic studies of GCs have shown diverse kinematics for red and blue GC populations. Often, the red GCs are found to follow the kinematics of the stellar spheroid of a galaxy (e.g. \citealt{Schuberth2010, Strader2011, Pota2013}), which might indicate a common formation history \citep{Shapiro2010}. Although the blue GC population often has a higher velocity dispersions (e.g. \citealt{Lee2008}), rotation signatures have been found for both the red and blue populations (e.g. \citealt{Arnold2011, Foster2011, Pota2013}). Studying higher moments of the GC LOS velocity distribution is challenging because it requires a large number of GC velocities and small uncertainties, but they can help to put constraints on the origin of a GC population (e.g. \citealt{Schuberth2010, Napolitano2014, Bianchini2017, Watkins2019}). 

Spectra are also required to obtain reliable metallicities because photometric estimates suffer from the age-metallicity degeneracy \citep{Worthey1994} and possible non-linearities in the colour-metallicity relation \citep{Richtler2006, Yoon2006}. Although GCs are usually old stellar objects, there are indications that the GC age distribution can spread over several Gyr \citep{Martocchia2018, Sesto2018, Usher2019}. 
While low signal-to-noise (S/N) spectra are sufficient to study LOS velocities, spectroscopic metallicities require higher quality data.

Fully exploiting the versatility of extragalactic GCSs requires large samples of spectroscopically studied GCs, but obtaining them is observationally time expensive with multi-object spectrographs. Nonetheless, extensive catalogues of GC velocities exist nowadays, for example in the core of the Fornax cluster \citep{Pota2018} or within the SAGES Legacy Unifying Globulars and Galaxies Survey (SLUGGS; \citealt{Brodie2014}) that has acquired radial velocities for more than 4000 GCs \citep{Pota2013, Forbes2017} and Calcium-triplet based metallicities for over 900 GCs \citep{Usher2012} in 27 early-type galaxies (ETGs).
These studies have shown that spectroscopic GC catalogues provide powerful tracers of the kinematics and metallicity out to $\leq 10\,R_\text{eff}$, far into the outer halo region of the host. However, with multi-object spectrographs, the inner regions of galaxies are usually avoided due to crowding of GCs in central regions and the difficulty of modelling the underlying bright galaxy light.

In this work, we used the high sensitivity in combination with a high spatial resolution and wide field of view (FOV) of the Multi Unit Spectroscopic Explorer (MUSE) instrument on the Very Large Telescope to extract a large number of background-cleaned GC spectra of 32 galaxies in the Fornax galaxy cluster. These galaxies were observed with MUSE as part of the Fornax3D survey (F3D, \citealt{F3D_Survey}).
F3D is a magnitude limited ($M_V \lesssim -17$ mag) survey that targets all massive galaxies within the virial radius of the Fornax cluster. 22 of these galaxies are ETGs, 10 are late-types galaxies (LTGs) with masses between 10$^{8}$ and 10$^{11}$ $M_\sun$, and for all galaxies, we could find at least one GC in the MUSE FOV. 

The F3D pointings cover between 2 -- 3 $R_\text{eff}$ of the host galaxies \citep{Iodice2019} and thus provide the opportunity to spectroscopically study GC systems in the inner parts of massive galaxies. Our catalogue provides a sample of LOS velocities of 722 GCs and metallicities of 238 GCs. In this work, we used this sample to test the utility of GCS as tracers of galaxy assembly by directly comparing their properties to the underlying host galaxies.
We compare the kinematics of the GC systems with that of their host galaxy, and investigate how the red and blue GCs trace the metallicity of the host galaxy. In an accompanying paper, we will focus more on the stellar population properties of the GCs themselves, and in particular we will infer the colour-metallicity relation from our sample of F3D GCs (\citealt{Fahrion2020c}, paper II in the following).

The paper is structured as follows: Sect. \ref{sect:data} gives an overview of the used MUSE data. In Sect. \ref{sect:Methods}, the methods for the spectra extraction and full spectral fitting are described, and the results are presented in Sect. \ref{sect:results}. We discuss our findings in Sect. \ref{sect:discussion} and give our conclusions in Sect. \ref{sect:conclusions}. In the Appendix \ref{app:catalogue}, we describe the F3D GC catalogue that is available online and give supplementary figures in App. \ref{app:all_the_plots}.

\begin{figure*}
\centering
\includegraphics[width=0.99\textwidth]{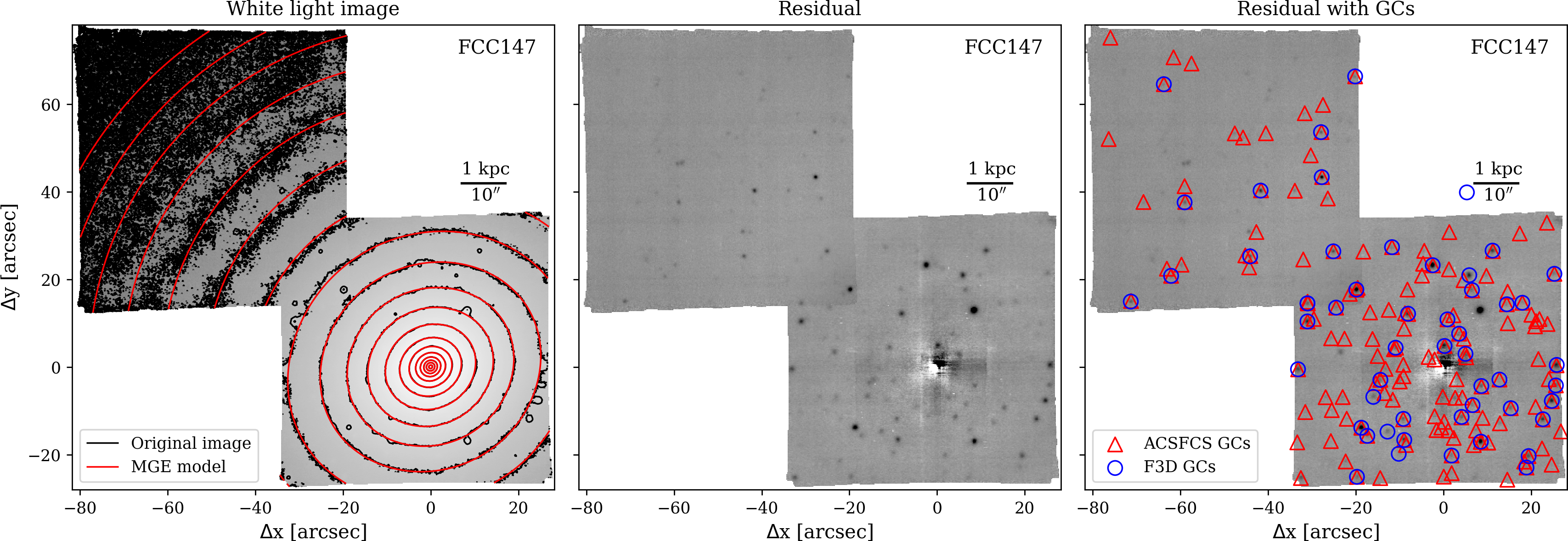}
\caption{Illustration of the MGE modelling used to create residual images in which GCs were detected. \textit{Left}: collapsed MUSE image of FCC\,147. Black and red contours illustrate isophotes of the original image and the MGE model, respectively. \textit{Middle}: residual image after subtracting the model from the image. Otherwise hidden point sources are now visible. \textit{Right}: Residual with GC highlighted. Red triangles show the position of GC candidates from the ACSFCS catalogue \citep{Jordan2015}. The blue circles show the position of confirmed F3D GCs in FCC\,147.}
\label{fig:MGE_example}
\end{figure*}

\section{MUSE data}
\label{sect:data}
F3D was carried out between July 2016 and December 2017 using MUSE \citep{Bacon2010} in the Wide Field Mode configuration that provides a 1$\times$1 arcmin$^2$ FOV per pointing, sampled at \mbox{0.2\arcsec $\times$ 0.2\arcsec}. The nominal wavelength range covers the optical from 4650\,\AA\,to 9300\,\AA\,with a sampling of 1.25\,\AA\,at a mean resolution of $\sim$ 2.5\,\AA.

Depending on individual galaxies, between one and three pointings with MUSE were acquired. The total integration times for central and middle pointings are $\sim$ 1 hour and halo pointings have a typical integration time of 1.5 hours, which was chosen to result in a limiting surface brightness of $\mu_B \approx 25$ mag arcsec$^{-2}$. The observations were carried out with a mean full width at half maximum (FWHM) of the point spread function (PSF) of $\sim$ 0.8\arcsec. This PSF in combination with the deep surface brightness level in the outer regions allow us to extract the spectra of a large number of point sources. 

The data reduction of the F3D data is described in detail in \cite{F3D_Survey} and \cite{Iodice2019}, and was performed with the MUSE pipeline 2.2 \citep{Weilbacher2012, Weilbacher2016}. It includes bias and flat-field correction, astrometric calibration, sky subtraction using dedicated sky observations, wavelength and flux calibration. To further reduce the contamination from sky lines, the Zurich Atmospheric Purge algorithm (ZAP; \citealt{Soto2016}) was applied. 

Maps of the LOS velocities were presented in \cite{Iodice2019} and in this work, we further used metallicity maps that were derived from the same line-strength measurements for a comparison to the GCs. More detailed maps of the stellar population properties will be subject of future work.
We also included FCC\,213 (NGC\,1399), the central galaxy of the Fornax cluster. For FCC\,213, we complemented the F3D MUSE pointings of the middle and outer parts with archival MUSE data of the central region (Prog. ID. 094.B-0903, PI: S. Zieleniewski).


\section{Methods}
\label{sect:Methods}
In the following, we describe our methods to detect the GCs in the MUSE data and to extract their spectra. We give details on how the LOS velocities and stellar population properties are determined from full spectral fitting. For a more detailed description, see \cite{Fahrion2019b}.

\subsection{Detection of GCs in the MUSE data}
At the distance of the Fornax cluster (20.9 Mpc \citealt{Blakeslee2009}), GCs appear as unresolved point sources in the collapsed MUSE images. The majority of GCs are hidden within the high surface brightness areas of their host galaxies and thus, the underlying light distribution of the galaxy has to be removed to detect these GCs. This was done by creating a Multi-Gaussian Expansion model (MGE, \citealt{Bendinelli1991, Monnet1992, Emsellem1994, Cappellari2002}) of every galaxy in our sample, which was subtracted from the image. In this way, a residual image was generated in which point sources such as GCs can be detected. To exploit the large wavelength coverage of MUSE, we did not use a single collapsed image for the MGE modelling, but instead cut the full MUSE cube into slabs of 500 wavelength slices (625 \AA) that are combined to a total of seven collapsed images from 4700 to 9000 \AA. The combination of several wavelength slices helps to improve the spatial S/N of the GCs and using seven instead of a single combined image further helps to reduce the contamination from emission line objects such as background star forming galaxies and planetary nebulae. In regions that have a visible dust feature, for example in the centre of FCC\,167 (see \citealt{Viaene2019}), we did not extract GCs.

Figure \ref{fig:MGE_example} shows an example of the MGE modelling for the ETG FCC\,147. In the residual image (middle panel), many point sources are clearly visible. We used \textsc{DAOStarFinder}, a Python implementation of the DAOFIND algorithm \citep{Stetson1987}, to detect those point sources in the image and to build the inital sample of GC candidates. Because most of the F3D galaxies were also covered by the ACS Fornax Cluster Survey (ACSFCS; \citealt{Jordan2007}), we used the catalogue of GC candidates from \cite{Jordan2015} for cross reference to remove the majority of background galaxies, foreground stars, and image artefacts from the GC sample (red triangles in Fig. \ref{fig:MGE_example}). The remaining contamination in this sample is very low ($\sim$ 1 \%), but although the ACSFCS catalogues are extensive and deeper than the MUSE data, they sometimes miss a small number of GCs. Therefore, we manually inspected the spectra of bright point sources that were not included in the ACSFCS catalogues. While redshifted background galaxies and dwarf stars are easily identified with their spectra, any additional GC candidate was checked for its LOS velocity before adding it to the catalogue. Per galaxy, there were usually only a few additional GCs found. Often, these lie in regions where the subtraction of the MGE model left a strong residual, for example in the disc of the S0 galaxy FCC\,170. 

FCC\,113, FCC\,161, FCC\,176, FCC\,179, FCC\,263, FCC\,285, FCC\,290, FCC\,306, FCC\,308, and FCC\,312 have no available ACSFCS GC catalogue and thus required that spectra of the full initial sample of point sources was checked to remove background galaxies and foreground stars. Because some of these galaxies also actively form stars (e.g. FCC\,312), HII regions were also among the point source sample and were filtered out. While background galaxies, bright foreground stars and line-emission objects such as HII regions can be easily found by visual inspection of the spectrum, the GC sample was finally cleaned after measuring the LOS velocity to confirm membership to the Fornax cluster (500 $<$ $v_\text{LOS} < 2500$ km s$^{-1}$). This range was based on the observed radial velocities of the F3D galaxies \citep{Iodice2019}, but the final sample only contains GCs with 800 $<$ $v_\text{LOS} < 2300$ km s$^{-1}$. The sources shown by blue circles in the right panel of Fig. \ref{fig:MGE_example} represent the final sample of confirmed GCs in FCC\,147.

Because of the velocity information and the central positioning of the pointings, the association of GCs to their host was rather straightforward. For most galaxies, all the GCs in the FOV could be associated to the observed host, however, we found four GCs in the pointing of FCC\,219 that appear to be associated to FCC\,213 because they have their velocities \mbox{$\sim 1300$ km s$^{-1}$}, close to the systemic velocity of FCC\,213, while the other GCs show velocities of \mbox{1700 - 2000 km s$^{-1}$}. These GCs might classify as intra-cluster GCs, but could also be bound to FCC\,213 (see, e.g. \citealt{Schuberth2008})
In FCC\,148, we found three GCs in the pointing that are likely associated to the nearby massive galaxy FCC\,147 due to their LOS velocities. In FCC\,184, three GCs were found that show a velocity difference to the host systemic velocity of $\sim$ 400 km s$^{-1}$, which might be possible intra-cluster GCs and one such GC was found in FCC\,182.

\subsection{Extraction of spectra}
We extracted the spectrum of each GC candidate from the MUSE cubes using a PSF-weighted circular aperture assuming a Gaussian shape with FWHM of 0.8\arcsec. The PSF weighting optimises the flux contribution from the GC, but especially in the high surface \mbox{brightness} area of the galaxy, it is essential to subtract the galaxy contribution to the spectrum. We therefore used an annulus aperture around each GC with an inner radius of 8 pixel and an outer radius of 13 pixel, centred on the GC position. To prevent contribution from neighbouring GCs to the background spectrum, we masked the positions of all GCs when extracting the background spectrum. 

For GCs with small galactocentric distances to their host's centre, the extraction of the background spectrum is particularly challenging due to the strongly varying surface brightness profile of the galaxy. For this reason, we extracted the background spectrum of close GCs ($<$ 10\arcsec) with a smaller annulus with inner and outer radii of 5 and 8 pixel, respectively. Still, the spectra of these GCs can be contaminated by galaxy light, especially in massive hosts. We found that a possible contamination mostly affects the GC metallicities that are then biased to higher values.

The spectral S/N was determined in a continuum region around 6500 \AA\,using the \textsc{estimateSNR} function of \textsc{PyAstronomy} \citep{pya}. For GCs with a S/N $\geq$ 3 \AA$^{-1}$, we determined LOS velocities and for GCs with S/N $\geq$ 8 \AA$^{-1}$, we additionally fitted for metallicities. GC candidates with S/N $<$ 3 \AA$^{-1}$ are removed from the final sample as we cannot confirm their nature or membership to a host galaxy. These S/N cuts are based on testing with single stellar population (SSP) model spectra that showed the stable recovery of metallicities down to S/N $\geq$ 8 \AA$^{-1}$ (similar to Appendix of \citealt{Fahrion2019b}).

\begin{figure}
\centering
\includegraphics[width=0.49\textwidth]{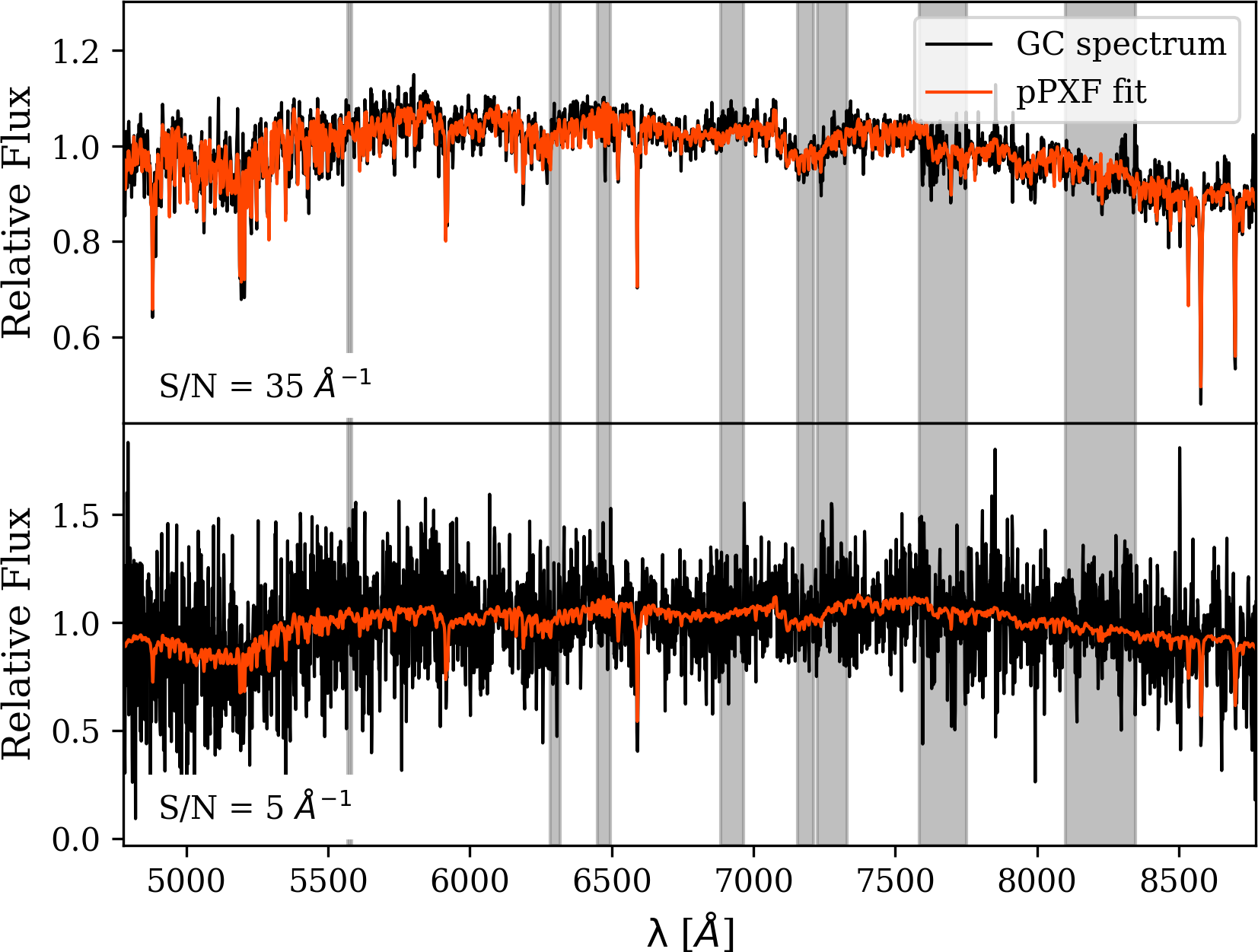}
\caption{Example of two GC spectra with \mbox{S/N $\sim$ 35 \AA$^{-1}$} (\textit{top}) and \mbox{S/N $\sim$ 5 \AA$^{-1}$} (\textit{bottom}). The original spectra are shown in black, the \textsc{pPXF} fit is shown in red. Regions with strong sky residual lines were masked from the fit (grey shaded areas). Both GCs were found in FCC\,161.}
\label{fig:spec_example}
\end{figure}

\begin{figure}
\centering
\includegraphics[width=0.49\textwidth]{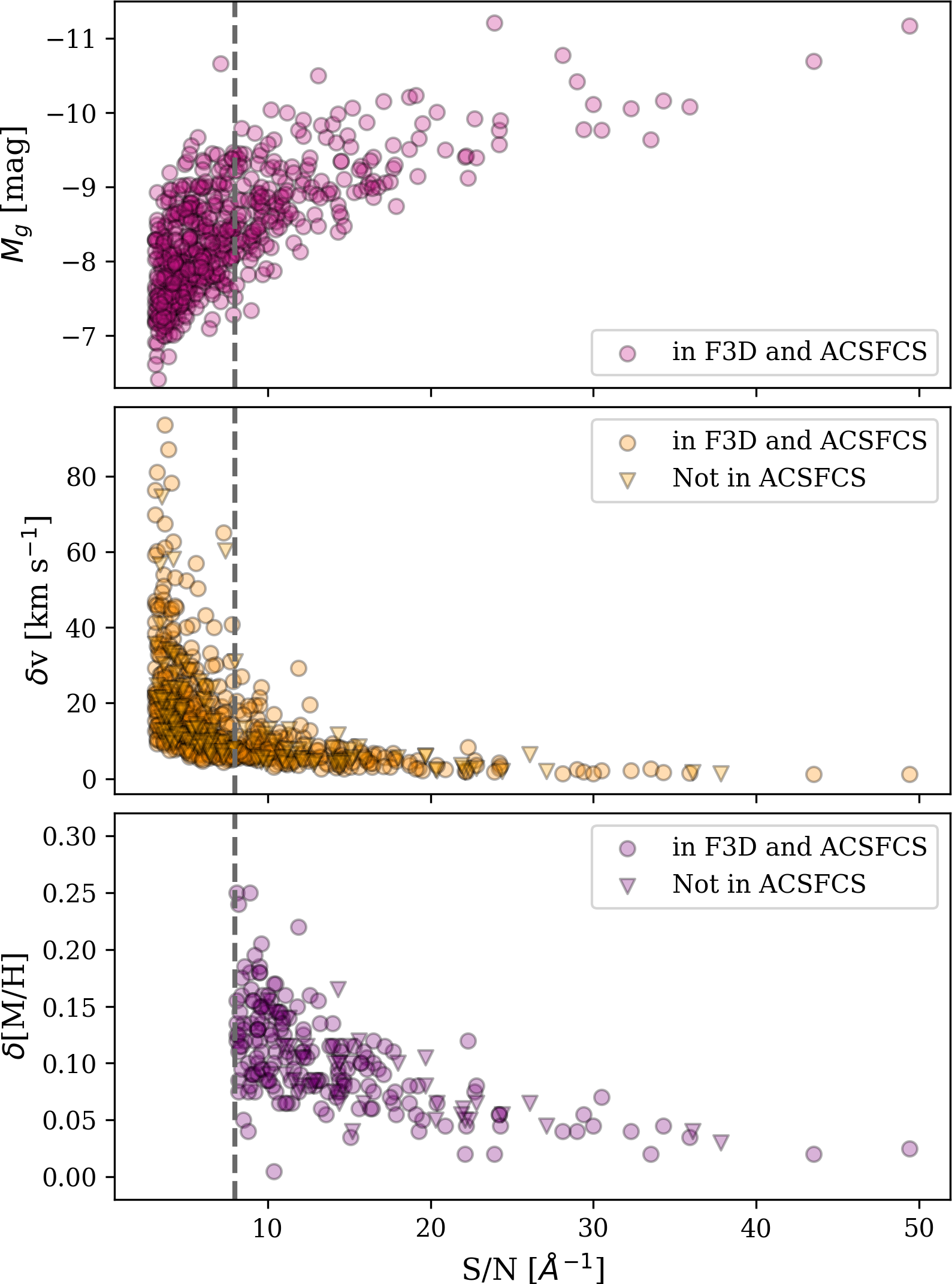}
\caption{Relation between GC S/N and absolute $g$-band magnitude from the ACSFCS catalogue (\textit{top}), LOS velocity uncertainty (\textit{middle}) and metallicity uncertainty (\textit{bottom}) for the sample of F3D GCs. The vertical line marks \mbox{S/N = 8 \AA$^{-1}$}, our limit for the metallicity measurement. In the top panel, we only show GCs with $M_g$ from the ACSFCS while the middle and bottom panel also include GCs that were not included in the catalogue of \cite{Jordan2015}.}
\label{fig:errors_vs_SNR}
\end{figure}

\begin{figure}
\centering
\includegraphics[width=0.49\textwidth]{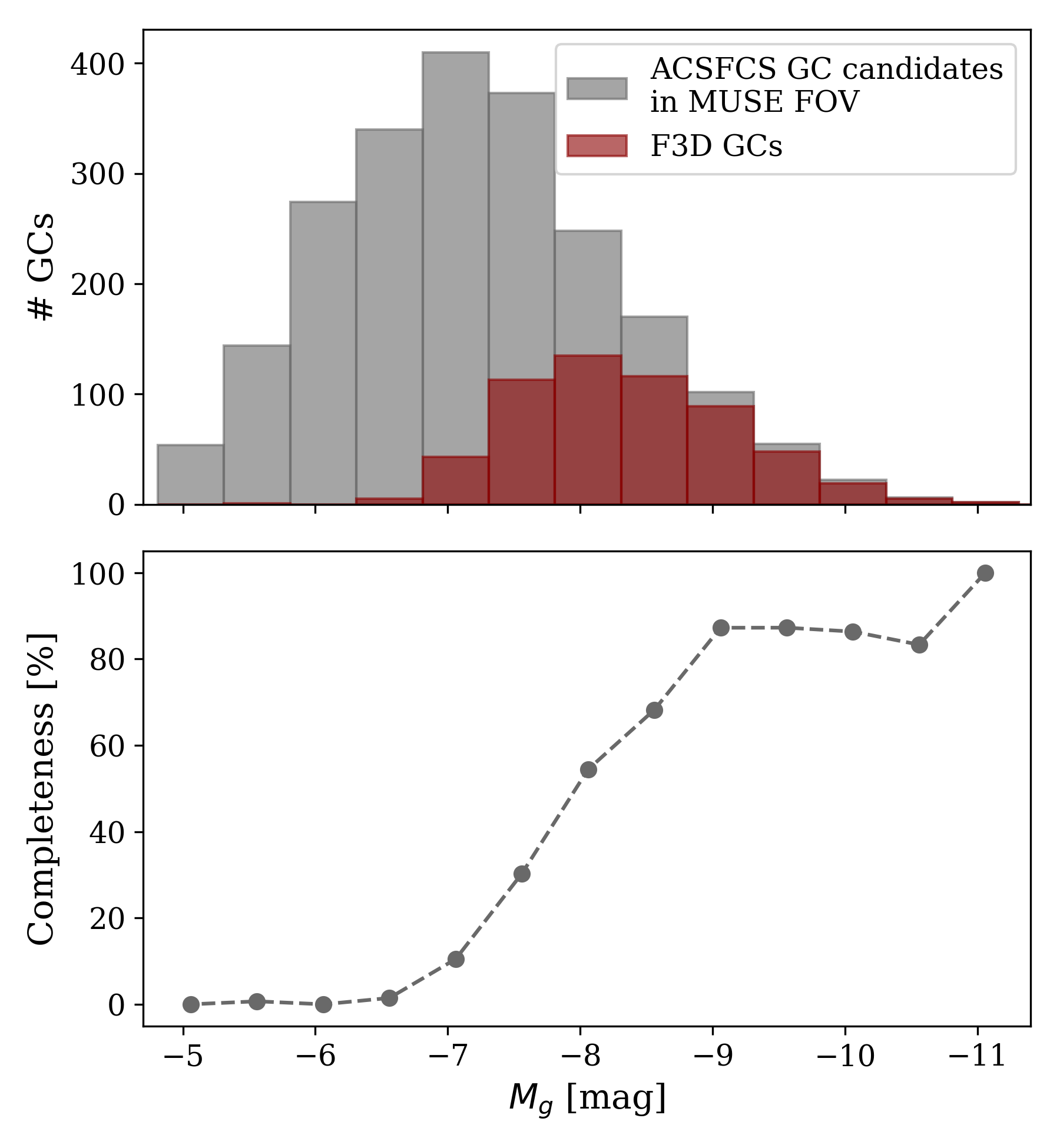}
\caption{Completeness of the F3D GC catalogue with respect to the ACSFCS catalogue \citep{Jordan2015}. \textit{Top:} Histogram of absolute $g$-band magnitudes. The grey bins show all ACSFCS GC candidates located in the MUSE FOVs of F3D. Those, that are confirmed GCs and are in our F3D catalogue are shown by the red bins. \textit{Bottom:} Completeness as ratio of number of GCs in F3D and all ACSFCS candidates as function of total $g$-band magnitude.}
\label{fig:completeness}
\end{figure}

\subsection{Full spectral fitting}
\label{sect:ppxf_fitting}
We fitted the GC spectra using full spectrum fitting with the penalised Pixel-fitting (\textsc{pPXF}) method \citep{Cappellari2004, Cappellari2017}.
\textsc{pPXF} is a full spectral fitting code that uses a penalised maximum likelihood approach to fit galaxy spectra with a combination of input template spectra. We used the SSP model spectra from the extended Medium resolution INT Library of Empirical Spectra (E-MILES, \citealt{Vazdekis2010, Vazdekis2016}) because of their broad wavelength coverage from \mbox{1680 to 50000 \AA}.
The E-MILES models with BaSTi isochrones \citep{Pietrinferni2004, Pietrinferni2006} give a grid of ages and total metallicities [M/H] between 30 Myr and 14 Gyr and [M/H] = $-2.27$ dex and $+0.04$ dex, respectively. Only so-called baseFe models are available in E-MILES that are based on empirical spectra and thus inherit the abundance pattern of the used stars. They have [Fe/H] = [M/H] at higher metallicities, but include $\alpha$-enhanced spectra at lowest metallicities. In paper II, we discuss the choice of SSP models further.

We used a MW-like double power-law (bimodal) inital mass function (IMF) with a high mass slope of 1.30 \citep{Vazdekis1996}. The model spectra have a spectral resolution of 2.51\,\AA\, in the wavelength region we used \citep{FalconBarroso2011}, approximately corresponding to the mean instrumental resolution of MUSE ($\sim 2.5\,\AA)$. We used the description of the line spread function from \cite{Guerou2016}, but we did not attempt to measure velocity dispersions because the intrinsic velocity dispersions of GCs (usually $< 20$ km s$^{-1}$) are below the spectral resolution of MUSE (\mbox{$\sim 80$ km s$^{-1}$}). 

Figure \ref{fig:spec_example} shows the spectra of two GCs of FCC\,161 and the corresponding \textsc{pPXF} fits as examples. Regions with residual sky or telluric lines were masked from the fit and are shown in grey.
We first fitted each GC with S/N $\geq$ 3 \AA$^{-1}$ for the LOS velocity with additive polynomials of degree 12 and no multiplicative polynomials. For GCs with S/N $\geq$ 8 \AA$^{-1}$, we fitted for the metallicity in a second step by keeping the LOS velocity fixed and using multiplicative polynomials of degree eight.
When fitting for stellar population properties such as age and metallicity, \textsc{pPXF} returns the weights of SSP models used to create the best fit. When regularisation is applied, this enables the construction of smooth distributions, for example of star formation histories (e.g. \citealt{Pinna2019, Boecker2020}). At the low S/N of the GC spectra, we did not use regularisation, but this should not affect the weighted mean metallicity. To limit effects from the well known age-metallicity degeneracy, we restricted the library to ages $\geq$ 8 Gyr.  As will be discussed paper II, this choice appears reasonable because most GCs have ages $>$ 8 Gyr with only very few exceptions. 

To estimate realistic uncertainties, we fitted each GC in a Monte Carlo-like approach (see also  \citealt{Cappellari2004, Wegner2012, Pinna2019, Bittner2019}). After the first fit using the original spectrum, we created 100 realisations of the spectrum by perturbing the noise-free best-fit spectrum with random draws in each wavelength bin from the residual (best-fit subtracted from original spectrum). The fit is then repeated and the LOS velocity - and if applicable - the metallicity were determined from the mean of the resulting distribution. The random uncertainty is given by the standard deviation assuming a Gaussian distribution.

The uncertainties on the velocity and metallicity depend on the S/N of the GC, as shown in Fig. \ref{fig:errors_vs_SNR}. In this figure, we plot the distribution of uncertainties in dependence of S/N. The uncertainty on the velocity is typically $<$ 20 km s$^{-1}$, but for GCs with S/N $\sim$ 3 \AA$^{-1}$, it can rise to values $>$ 50 km s$^{-1}$. Typical metallicity uncertainties are $\sim 0.15$ dex. We also show the relation between spectral S/N and absolute $g$-band magnitude from the ACSFCS catalogue assuming a distance to the Fornax cluster of 20.9 Mpc \citep{Blakeslee2009}. In general, the brighter GCs have higher S/N, but there is a large scatter in the relation. This scatter is not surprising because the GC spectra have different total exposure times depending on their location in the MUSE mosaics. Their S/N depends also on the contrast with the underlying galaxy background and the absence of strong absorption lines at low metallicities increases uncertainties for metal-poor GCs.

\begin{table*}
\centering
\caption{Overview of F3D galaxies and the number of extracted GCs. The galaxies are ordered by their FCC number. Masses are from \cite{Iodice2019, Liu2019}.}
\label{tab:F3D_GCs}
\begin{tabular}{*{11}{c}}\hline
FCC & Altern. name &  RA & DEC & log($M_\ast$) & $R_\text{eff}$ & Type & $N_\text{S/N > 3}$ & $N_\text{S/N > 8}$  & $<R_\text{GCs}>$ & $<$[M/H]$_\text{GCs}>$ \\ 
&	& (J2000) & (J2000) & log($M_\sun$) & (\arcsec) & & & & ($R_\text{eff}$) & (dex) \\
(1) & (2) & (3) & (4) & (5) & (6) & (7) & (8) & (9) & (10) & (11) \\
\hline\hline
083 & NGC\,1351 & 03:30:35.1 & $-$34:51:14  & 10.5 & 35.7 & E5 & 57 & 25 & 0.70 &   $-$0.75 $\pm$ 0.49 \\
090 & PGC\,13058 & 03:31:08.1 & $-$36:17:19 & 8.9 & 12.1 &  E4 & 4 & 1 & 1.11&  $-$1.99 \\
113 & ESO\,358-015 & 03:33:06.8 & $-$34:48:26 & 8.3 & 20.6 & Scd & 4 & 2 & 0.75 & $-$1.58 $\pm$ 0.34  \\
119 & -- & 03:33:33.7 & $-$33:34:18 & 9.0 & 17.4 & S0 & 1 & 0 & 1.05  & -- \\
143 & NGC\,1373 & 03:34:59.1 & $-$35:10:10 & 9.4 & 11.0 & E3 & 20 & 4 &  1.91 & $-$1.24 $\pm$ 0.21 \\
147 & NGC\,1374 & 03:35:16.8 & $-$35:13:34 & 10.4 & 24.8 &  E0 & 55 & 20 & 1.20 & $-$0.81 $\pm$ 0.57 \\
148 & NGC\,1375 & 03:35:16.8 & $-$35:15:56 & 9.8 & 28.3 & S0 &  2 & 2 &  2.10 & $-$1.85 $\pm$ 0.14  \\
153 & IC\,1963 & 03:35:30.9 & $-$34:26:45 & 9.9 & 19.8 & S0 & 14 & 2 & 1.26 & $-$1.52 $\pm$ 0.08 \\
161 & NGC\,1379 & 03:36:04.0 & $-$35:26:30 & 10.4 & 28.6 & E0 & 71 & 32 & 0.97 & $-$0.70 $\pm$ 0.44 \\
167 & NGC\,1380 & 03:36:27.5 & $-$34:58:31 & 11.0 & 56.4 & S0 & 59 & 17 & 0.68 & $-$0.32 $\pm$ 0.56 \\
170 & NGC\,1381 & 03:36:31.6 & $-$35:17:43 & 10.4 & 15.9 & S0 & 22 & 9 &  2.00 & $-$0.74 $\pm$ 0.51 \\
176 & NGC\,1369 & 03:36:45.0 & $-$36:15:17 & 9.8 & 53.7 &SB & 11 & 1 & 0.41 & $-$2.05 \\
177 & NGC\,1380A & 03:36:47.4 & $-$34:44:17 & 9.9 & 35.9 & S0 & 19 & 8 &  0.68 & $-$1.23 $\pm$ 0.42 \\
179 & NGC\,1386 & 03:36:46.3 & $-$35:59:57 & 10.2 & 30.0 &  Sa & 2 & 0  & 0.96 &  -- \\
182 & -- & 03:36:54.3 & $-$35:22:23 & 9.2 & 9.9 & SB0 & 9 & 3  & 1.74 & $-$0.81 $\pm$ 0.28  \\
184 & NGC\,1387 & 03:36:56.9 & $-$35:30:24 & 10.7 & 35.5 & SB0 & 49 & 18  & 1.25 & $-$0.20 $\pm$ 0.47 \\
190 & NGC\,1380B & 03:37:08.9 & $-$35:11:37 & 9.7 & 18.3 & SB0 & 32 & 12  & 1.22 &$-$1.38 $\pm$ 0.33 \\
193 & NGC\,1389 & 03:37:11.7 & $-$35:44:40 & 10.5 & 28.2 & SB0 & 16 & 1  &0.72 & $-$1.49  \\
213 & NGC\,1399 & 03:38:29.2 & $-$35:27:02 & 11.4 & 308 &  E1 & 111 & 26 & 0.12 & $-$0.50 $\pm$ 0.55 \\
219 & NGC\,1404 & 03:38:52.1 & $-$35:35:38 & 11.1 & 161.0 &  E2 & 22 & 4 & 0.22 & $-$0.13 $\pm$ 0.29 \\
249 & NGC\,1419 & 03:40:41.9 & $-$37:30:33 & 9.7 & 9.6 & E0  &  39 & 8 & 1.72 & $-$0.80 $\pm$ 0.44 \\
255 & ESO\,358-G50 & 03:41:03.4 & $-$33:46:38 & 9.7 & 13.8 & S0 & 19 & 6 &  1.38 &  $-$1.17 $\pm$ 0.49 \\
263 & ESO\,358-051 & 03:41:32.2 & $-$34:53:17 & 8.6 & 27.2 & SB & 1 & 0 & 0.61 & -- \\
276 & NGC\,1427 & 03:42:19.2 & $-$35:23:36 &10.3 & 44.7 &  E4 &  53 & 26 & 0.77 & $-$0.54 $\pm$ 0.45 \\
277 & NGC\,1428 & 03:42:22.6 & $-$35:09:10 & 9.5 & 12.8 &  E5 & 8 & 1  & 1.73 & $-$1.17 \\
285 & NGC\,1437A & 03:43:01.8 & $-$36:16:11 & 8.3 & 49.9 & Sd & 1 & 0  & 0.68 &  --  \\
290 & NGC\,1436 & 03:43:37.0 & $-$35:51:13 & 9.8 & 48.5 & Sc & 2 & 1  & 0.34 &  $-$1.22 \\
301 & ESO 358-G59 & 03:45:03.5 & $-$35:58:17 & 9.3 & 11.7 & E4 & 1 & 0  & 2.05 & -- \\
306 & -- & 03:45:45.3 & $-$36:20:40 & 8.0 & 9.7 & SB & 1 & 1 & 0.38 & $-1.13$ \\
308 & NGC\,1437B & 03:45:54.7 & $-$36:21:25 & 8.6 & 37.1 &  Sd & 7 & 3  & 1.04 & $-$1.57 $\pm$ 0.19  \\
310 & NGC\,1460 & 03:46:13.7 & $-$36:41:43 & 9.7 & 35.6 & SB0 & 13 & 4 &0.51 &  $-$0.82 $\pm$ 0.52 \\
312 & ESO\,358-063 & 03:46:18.9 & $-$34:56:31 & 10.2 & 109.5 &  Scd & 3 & 0  & 0.47&  --  \\\hline
\end{tabular}
\tablefoot{(1) Galaxy name from \cite{Ferguson1989} and (2) alternative name. (3) and (4): Right ascension and declination. (5) - (7): Stellar mass, effective radius and morphological type from \cite{Iodice2019, Iodice2019a} and \cite{Liu2019}. (8) and (9): Number of GCs with S/N $>$ 3 and $>$ 8 respectively. (10) Mean projected radius and (11) mean metallicity of the F3D GCs. We found four possible intra-cluster GCs, three in FCC\,184, one in FCC\,182.
}
\end{table*}

\section{Results}
\label{sect:results}
We briefly summarise our sample of GCs in F3D in the following. Then, the kinematic modelling of the GC systems is described and we compare GC metallicities to the metallicity profile of the host galaxies.

\subsection{Sample of globular clusters}
In total, we determined the LOS velocities for 722 GCs in 32 galaxies and the metallicities of 238 GCs. We could identify at least one GC in every F3D galaxy. Table \ref{tab:F3D_GCs} gives an overview of the extracted GCs per F3D galaxy and in App. \ref{app:all_the_plots}, we show several plots for each galaxy to illustrate the sample. 

From the galaxies that were also covered by the ACSFCS, we can estimate the completeness level of our GC extraction. Figure \ref{fig:completeness} shows the histogram of $g$-band magnitudes for all ACFCS GC candidates that are located within the MUSE pointings of F3D in comparison to those that are also in the final F3D GC catalogue. The lower panel shows the completeness as a function of magnitude, computed as the ratio between the number of GCs in F3D and ACSFCS, respectively. We reached a completeness of $\sim$ 50 \% at $M_g \sim -8$ mag. This is a conservative limit, because we applied no probability cut on the ACSFCS GCs. The faintest GCs in our sample have magnitudes of $M_g \approx -7$ mag ($g \approx$ 24.6 mag), corresponding to $\sim 10^{5} M_\sun$.  We separated the red and blue GC populations at a fixed colour of ($g - z$) = 1.16 mag (e.g. \citealt{Peng2008, Liu2019}). In the final sample, 60\% of the GCs are blue.

\begin{figure*}
\centering
\includegraphics[width=0.54\textwidth]{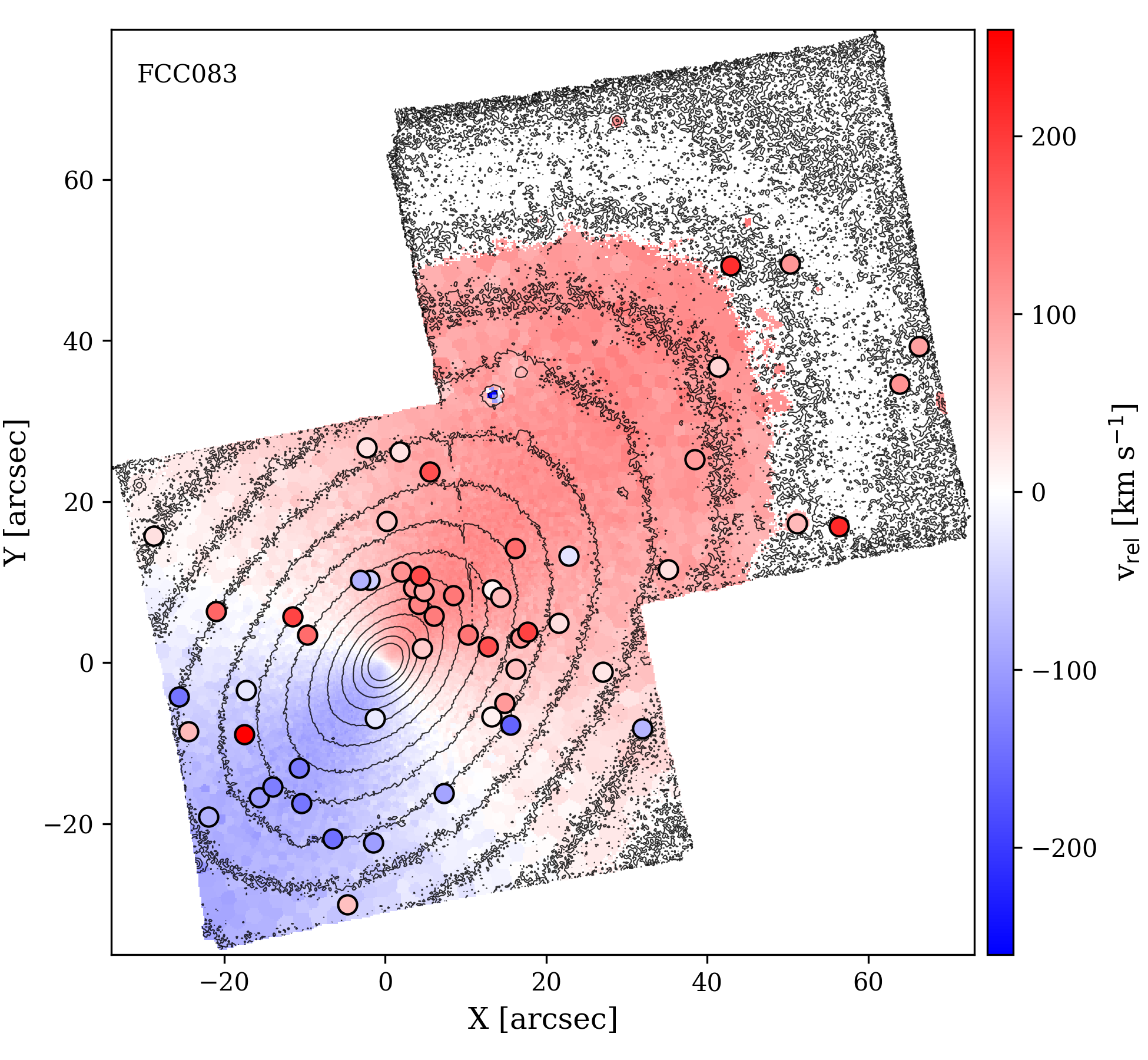}
\includegraphics[width=0.45\textwidth]{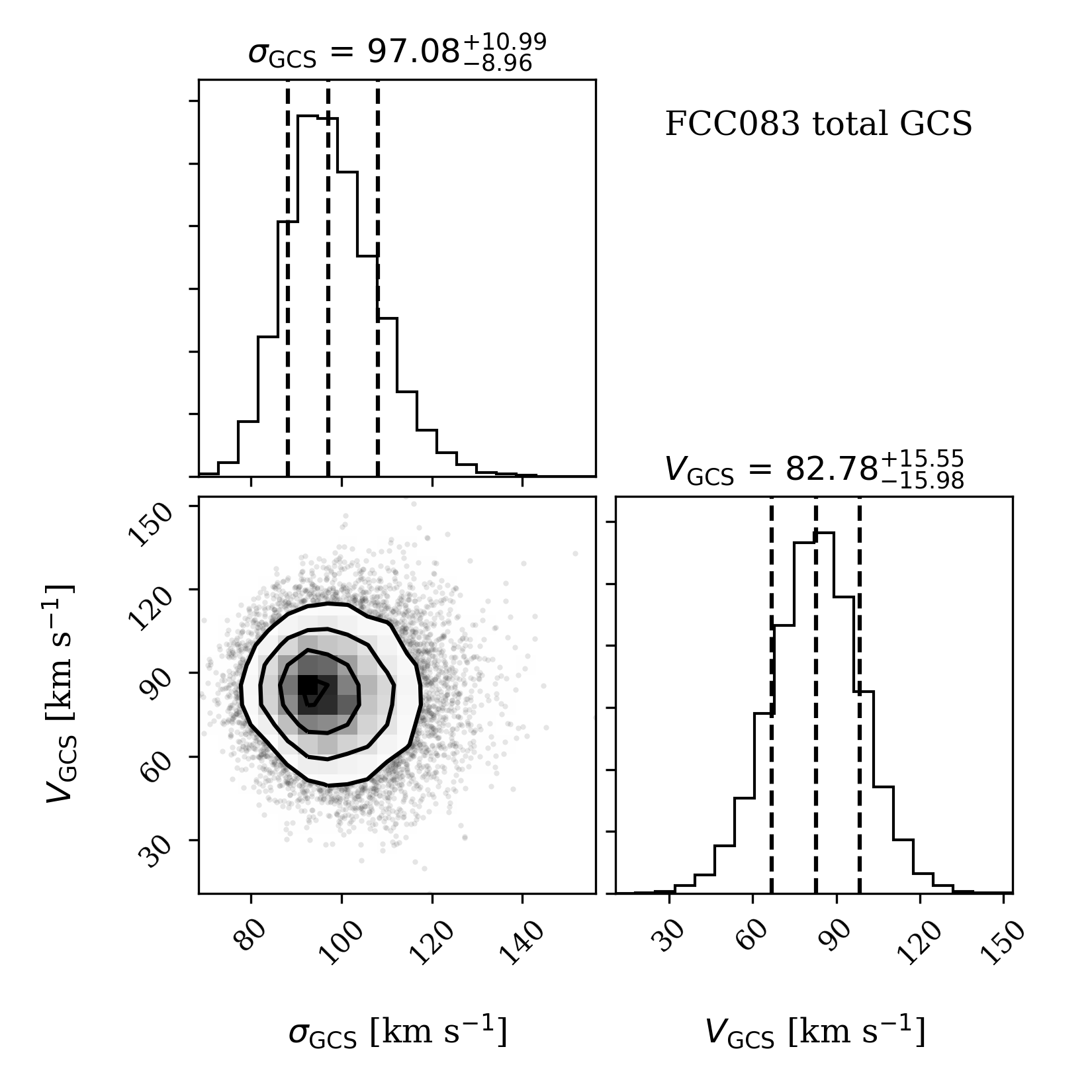}
\caption{Kinematic modelling of the GC system of FCC\,083. \textit{Left}: LOS velocity map of FCC\,083 with GC velocities overplotted as circles. We used the same colourbar scaling for the stellar light bins and the GCs. The black contours are isophotes to guide the eye. \textit{Right}: Posterior distributions for $V_\text{GCS}$ and $\sigma_\text{GCS}$ of the MCMC fit to the GCS of FCC\,083 (Eq. \ref{eq:kin_model}). For $\theta_0$, we used a Gaussian prior with $\theta_0$ = 142 $\pm$ 1$^\circ$ \citep{Iodice2019}.}
\label{fig:vel_maps_example}
\end{figure*}

The majority of the F3D GCs are hosted by ETGs, while the number of GCs in LTGs is low. In the ETGs, we found 686 GCs in 22 galaxies, corresponding to 31 GCs per galaxy on average. We could identify 36 GCs in the 10 LTGs of F3D. FCC\,176 is the only LTG in our sample for which we could identify more than 10 GCs in the FOV.
The low number of detected GCs in the LTGs might have different reasons. Firstly, the LTGs have lower masses than the ETGs in the inner region of Fornax and thus the number of expected GCs is lower because the total mass of a GC system depends on the mass of the host galaxy \citep{Forbes2018, Liu2019} and LTGs have lower specific frequencies than ETGs in general \citep{Georgiev2010}. Secondly, most of the LTGs in our sample are actively star forming and have irregular morphologies, which made the detection of GCs in the FOV challenging because of strong residuals from MGE modelling. Thirdly, most of the LTGs were covered by a single pointing and thus we have no access to GCs at larger radii. Lastly, the LTGs were not covered by the ACSFCS and thus no catalogue of GC candidates is available that would help to identify possible GCs.  

In the Scd galaxy FCC\,113, we were able to detect four GCs, one of them appears to be located in the photometric centre of the galaxy and thus could classify as the nuclear star cluster of this galaxy. In addition, in FCC\,290 we found a star cluster with a synthetic MUSE colour of ($g - z$) $\sim$ 0.3 mag. For this blue star cluster, we found a stellar age of $\sim$ 2 Gyr, significantly younger than the other GCs in our sample. This star cluster might therefore be a genuine young star cluster, maybe similar to those that are found in star forming galaxies (e.g. \citealt{LarsenRichtler1999, Adamo2010, Fedotov2011}). However, a higher S/N would be required to confirm this young age spectroscopically. Although our sample of GCs in LTGs is relatively small, these are the first spectroscopically confirmed GCs of LTGs in Fornax and might provide a basis for follow-up studies. 

\begin{figure*}
\centering
\includegraphics[width=0.99\textwidth]{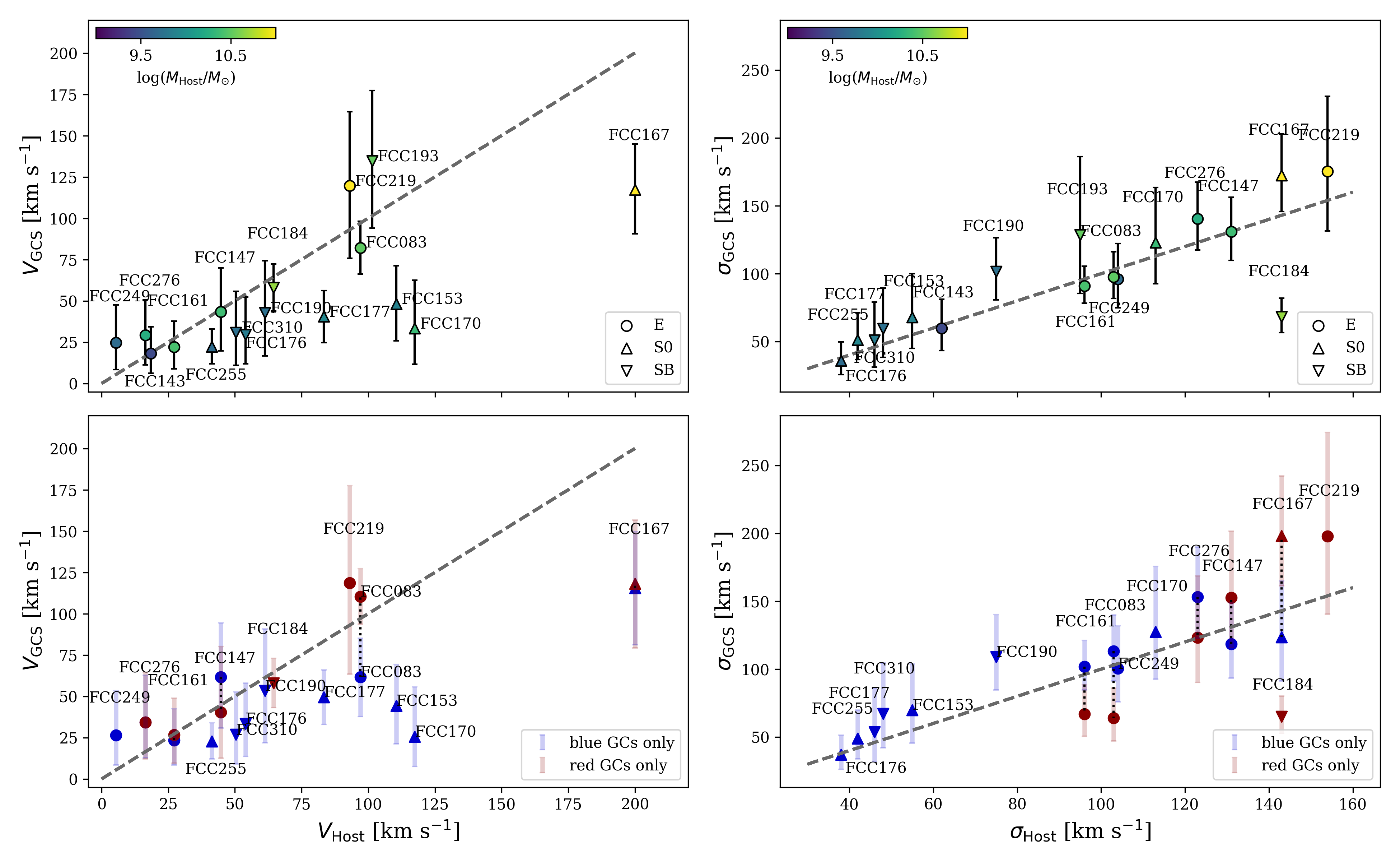}
\caption{GC rotation velocities (\textit{left}) and velocity dispersions (\textit{right}) compared to the stellar body of the galaxies. The stellar rotation amplitudes and velocity dispersions were extracted from the kinematic maps at 1 $R_\text{eff}$ where possible. In the upper panels, $V_\text{GCS}$ and $\sigma_\text{GCS}$ for the total GCS are shown. The colour gives the stellar mass of the host galaxy and the symbols show the galaxy morphology. The bottom panel shows the values, when only the red or blue GCs were modelled. For galaxies with more than 10 GCs in either population, a dotted line connects the values. The dashed line in all panels gives the one-to-one relation. FCC\,213 is not shown because no kinematic map was available.}
\label{fig:V_and_sig}
\end{figure*}

\subsection{Rotation of globular cluster systems}
\label{sect:rotation}
The extracted GC LOS velocities enable a comparison to their host galaxies using the kinematic maps that were presented in \cite{Iodice2019}. As one example, Fig. \ref{fig:vel_maps_example} shows the stellar LOS velocity map of FCC\,083, and we overplotted the LOS velocities of the GCs with circles. Because \cite{Iodice2019} used a slightly different setup to extract the LOS velocity, for example with a different masking of sky regions and a different wavelength range, we corrected any possible systematic offsets in velocity between the map and the GCs by separately fitting the central pixel of each galaxy with the same \textsc{pPXF} setup that was used for the GCs. We found no offset \mbox{$> 20$ km s$^{-1}$}.

The GC system of FCC\,083 clearly shows rotation along the same axis as the galaxy and we found similar behaviours in other galaxies such as FCC\,147, FCC\,184, FCC\,190, FCC\,153, FCC\,167, FCC\,193, and FCC\,219 as can be clearly seen in App. \ref{app:all_the_plots}. In these figures, we show the LOS velocities of the GCs in comparison to the stellar light, both on the LOS velocity maps and as a profile along the main kinematic axis. We also found several galaxies that do not show any sign of rotation in their GC system, for example FCC\,161 and FCC\,213. These two galaxies also show very low rotation amplitudes in their stellar body. 

\begin{table*}
\centering
\caption{Rotation amplitude $V_\text{GCS}$ and velocity dispersion $\sigma_\text{GCS}$ for the GC systems of the F3D galaxies.}
\label{tab:F3D_GCs_kin}
'
\begin{tabular}{*{10}{c}}\hline
Galaxy & $\sigma_\text{e}$ & $N_\text{red}$ & $N_\text{blue}$ & $V_\text{GCS}$ & $\sigma_\text{GCS}$ & $V_\text{red, GCS}$ & $\sigma_\text{red, GCS}$ & $V_\text{blue, GCS}$ & $\sigma_\text{blue GCS}$    \\ 
&	(km s$^{-1}$) & & & (km s$^{-1}$) & (km s$^{-1}$) & (km s$^{-1}$) & (km s$^{-1}$) & (km s$^{-1}$) & (km s$^{-1}$) \\ 
(1) & (2) & (3) & (4) & (5) & (6) & (7) & (8) & (9) & (10) \\
\hline\hline 
FCC083 & 103 & 20 & 37 &  82.8$^{+15.6}_{-16.0}$ & 97.1$^{+11.0}_{-9.0}$ &110.3$^{+16.9}_{-16.8}$ & 64.0$^{+12.7}_{-9.7}$ & 61.6$^{+23.6}_{-23.8}$ & 113.4$^{+15.4}_{-12.8}$ \\
FCC143 & 62 &  0 & 20 & 18.2$^{+16.0}_{-11.9}$ & 59.7$^{+12.4}_{-9.4}$ & -- & -- & -- & -- \\
FCC147 & 131 &  22 & 30 & 43.4$^{+26.5}_{-23.8}$ & 131.1$^{+14.6}_{-12.3}$ & 40.3$^{+39.7}_{-27.8}$ & 152.6$^{+28.3}_{-21.2}$ &61.6$^{+32.9}_{-30.8}$ & 118.6$^{+17.6}_{-14.6}$ \\
FCC153 & 55 & 1 & 13 & 48.2$^{+23.0}_{-22.4}$ & 68.0$^{+18.5}_{-13.2}$ & -- & -- & 44.3$^{+25.0}_{-23.0}$ & 69.9$^{+19.7}_{-14.1}$ \\
FCC161 & 96 & 22 & 48 & 22.1$^{+15.6}_{-13.3}$ & 90.9$^{+8.4}_{-7.3}$ & 26.7$^{+22.1}_{-17.3}$ & 67.0$^{+12.4}_{-9.4}$ & 23.5$^{+19.1}_{-15.1}$ & 101.7$^{+11.2}_{-9.9}$ \\
FCC167 & 143 & 38 & 21 & 117.4$^{+27.5}_{-26.8}$ & 172.4$^{+17.5}_{-15.4}$ &118.2$^{+38.5}_{-39.0}$ & 198.2$^{+25.4}_{-21.3}$ & 115.6$^{+34.1}_{-34.5}$ & 123.5$^{+24.0}_{-18.5}$ \\
FCC170 & 113 & 4 & 18 & 33.3$^{+29.1}_{-21.7}$ & 122.9$^{+23.4}_{-17.5}$ & -- & -- & 25.6$^{+30.1}_{-18.1}$ & 127.5$^{+27.8}_{-20.1}$ \\
FCC176 & 43 & 1 & 10 & 29.5$^{+22.7}_{-17.6}$ & 51.0$^{+16.3}_{-11.4}$ & -- & -- & 33.4$^{+24.5}_{-19.9}$ & 53.8$^{+18.6}_{-12.6}$ \\
FCC177 & 42 & 3 & 16 & 40.5$^{+15.7}_{-15.8}$ & 51.4$^{+11.4}_{-8.5}$ & -- & -- & 49.5$^{+16.5}_{-16.6}$ & 49.0$^{+12.0}_{-8.8}$ \\
FCC184 & 143 & 40 & 9 & 57.9$^{+14.4}_{-14.1}$ & 68.3$^{+7.9}_{-6.8}$ & 57.9$^{+15.1}_{-14.6}$ & $65.0^{+8.8}_{-7.2}$ & -- & -- \\
FCC190 & 75 & 5 & 27 & 42.7$^{+31.6}_{-26.0}$ & 101.6$^{+14.4}_{-12.0}$ & -- & -- & 53.3$^{+37.5}_{-31.6}$ & 109.1$^{+17.9}_{-14.1}$ \\
FCC193 & 95 & 7 & 9 & 134.6$^{+42.8}_{-40.6}$ & 128.5$^{+33.3}_{-24.8}$ & -- & -- &-- & -- \\
FCC213 & -- & 68 & 41 & 51.8$^{+27.2}_{-25.9}$ & 228.8$^{+16.3}_{-14.6}$ &30.3$^{+25.6}_{-19.8}$ & 192.5$^{+18.2}_{-15.8}$ & 102.1$^{+55.9}_{-51.7}$ & 282.0$^{+35.0}_{-29.6}$ \\
FCC219 & 154 & 17 & 7 & 119.6$^{+44.8}_{-43.9}$ & 175.5$^{+32.0}_{-25.4}$ & 118.5$^{+59.0}_{-55.0}$ & 197.9$^{+44.2}_{-33.2}$ &-- & -- \\
FCC249 & 104 & 5 & 24 & 24.7$^{+22.8}_{-16.4}$ & 96.0$^{+15.2}_{-12.2}$ & -- & -- & 26.4$^{+26.9}_{-18.0}$ & 100.4$^{+18.3}_{-14.1}$ \\
FCC255 & 38 & 1 & 18 & 22.2$^{+10.6}_{-10.3}$ & 36.0$^{+7.9}_{-6.0}$ & -- & -- & 22.8$^{+11.2}_{-10.8}$ & 37.1$^{+8.3}_{-6.3}$ \\
FCC276 & 123 & 18 & 35 & 29.2$^{+21.3}_{-18.0}$ & 140.5$^{+15.5}_{-13.3}$ & 34.3$^{+30.3}_{-22.3}$ & 123.4$^{+26.2}_{-19.2}$ &34.3$^{+28.5}_{-21.7}$ & 153.2$^{+21.2}_{-17.5}$ \\
FCC310 & 48 & 2 & 11 & 30.8$^{+25.0}_{-19.8}$ & 59.5$^{+17.3}_{-12.1}$ & -- & -- & 26.9$^{+25.9}_{-17.9}$ & 67.1$^{+21.7}_{-14.4}$ \\ \hline
\end{tabular}
\tablefoot{(1) Galaxy name. (2) average LOS velocity dispersion within 1 $R_\text{eff}$ from \cite{Iodice2019}. (3), (4): number of red and blue GCs in the F3D sample. (6) - (10): rotation amplitude and velocity dispersion. If ten or more GCs are available in the red and blue populations, we determined $V_\text{GCS}$ and $\sigma_\text{GCS}$ for these separately.}
\end{table*}

\subsubsection{Kinematic modelling}
In order to quantify the rotational motion, we modelled the kinematics of the GC systems (GCSs) with a simple model following the description by \cite{Veljanoski2016}. 
The rotation amplitude $V_\text{GCS}$ of the GC system is described as \citep{Cote2001}:
\begin{equation}
v_{\text{GC}, i}(\theta) = v_\text{0} + V_{\text{GCS}}\,\text{sin}(\theta_i - \theta_0),
\label{eq:v_vs_PA}
\end{equation}
where $v_{\text{GC}, i}$ is the LOS velocity of the $i$-th GC at position angle $\theta_i$ and $V_{\text{GCS}}$ then gives the rotation amplitude of the total GC, that reaches its maximum along position angle $\theta_0$ + 90$^\circ$. $v_\text{0}$ is the mean velocity of the GCS that could in principle deviate from the systemic velocity of the host galaxy, but for the sake of simplicity we fixed this parameter to the host galaxy's systemic velocity. This model also assumes radially invariant rotation velocities and velocity dispersions, and we did not correct for inclination. The reported values are projected quantities.

Under the assumption that the velocity dispersion $\sigma$ can be represented by a Gaussian, it is described as:
\begin{equation}
\sigma^2 = (\Delta v_\text{GC, i})^{2} + {\sigma}^{2},
\label{eq:vel_disp} 
\end{equation}
where $\Delta v_\text{GC, i}$ is the velocity uncertainty for the $i$-th GC and $\sigma$ denotes the LOS velocity dispersion. Combining Eq. \ref{eq:v_vs_PA} and \ref{eq:vel_disp}, allows us to construct a model with the likelihood described by:
\begin{equation}
\mathcal{L} = \prod \limits_{i} \frac{1}{\sqrt{2 \pi \sigma^2}} \text{exp} \left(- \frac{(v_\text{GC, i} - (v_0 + V_\text{GCS}\,\text{sin}(\theta_i - \theta_0)))^2}{2 \sigma^2} \right).
\label{eq:kin_model}
\end{equation}
In this model, the radial velocities $v_\text{GC, i}$ of the GCs, their uncertainties ($\Delta v_\text{GC, i}$), and the position angles $\theta_i$ are the input data. $V_\text{GCS}$, $\sigma_\text{GCS}$ and $\theta_0$ are free parameters of the model. We implemented Eq. \ref{eq:kin_model} in \textsc{emcee} \citep{emcee}, a python implementation of the Markov chain Monte Carlo (MCMC) sampler to sample the posterior probability distribution function. We used flat positive priors for $V_\text{GCS}$ and $\sigma_\text{GCS}$, but used a Gaussian prior for the position angle $\theta_0$ given by the host's kinematic position angle and uncertainty \citep{Iodice2019}. This is necessary because the GC sample is geometrically limited, often along a preferred axis, due to the positioning of the MUSE pointings. In the right panel of Fig. \ref{fig:vel_maps_example}, we show the resulting posterior distributions for FCC\,083 using the described priors. However, FCC\,083 has a strongly rotating GCS and we found that also a flat prior on $\theta_0$ results in a well constrained distribution.

We modelled all galaxies with ten or more GCs, and for galaxies that have ten or more red or blue GCs, respectively, we modelled these populations separately by only including the GCs of the respective colour. For a few galaxies, we can model both the red and blue GC population separately. We list the resulting values for $V_\text{GCS}$ and $\sigma_\text{GCS}$ in Tab. \ref{tab:F3D_GCs_kin}. The uncertainties refer to the 16th and 84th percentile of the MCMC parameter distribution.

\subsubsection{Comparison to host galaxy}
We compare the rotation amplitude and velocity dispersion of the GCSs to the stellar bodies in Fig. \ref{fig:V_and_sig}. 
In the right panel, $\sigma_\text{GCS}$ is plotted against $\sigma_e$, the average velocity dispersion within one effective radius from \cite{Iodice2019}. Additionally, we extracted the velocity of the host galaxy along its kinematic major axis at the effective radius, where possible. For FCC\,176, FCC\,310, and FCC\,219, the MUSE FOV does not cover one effective radius and for those, we used the largest radius covered by the maps. In general, we found that the exact choice of the extraction radius does not significantly influence this comparison.

The velocity dispersion found in the GCs is in good agreement with that of the stars and increases with stellar mass, indicating that the GCs can be used as kinematic tracers of the enclosed mass. This correlation is also seen when the red and blue GCs are modelled separately, independent of the galaxy type. However, when only using blue GCs, the relation with the stellar velocity dispersion appears tighter.
FCC\,184 is an outlier from this relation with a low velocity dispersion in its GCS. FCC\,184 also stands out with having a large number of red, very metal-rich GCs (see Fig. \ref{fig:maps6}) that seem to follow the (low) rotation amplitude of the galaxy as the left panel of Fig. \ref{fig:V_and_sig} shows. This could indicate that these GCs have a common origin possibly in the disc of FCC\,184 and constitute a dynamically cold component in the galaxy.

The top left panel of Fig. \ref{fig:V_and_sig} compares the rotation amplitudes of GCS and stars. This comparison shows that the GCs trace the rotation closely in the elliptical galaxies, similar to what was found by \cite{Pota2013} for 12 ETGs. In contrast, the three edge-on S0 galaxies (FCC\,170, FCC\,153, and FCC\,177, \citealt{Pinna2019, Pinna2019b}) show low rotation amplitudes in the GCSs compared to that found in the stars. This might be caused by the strongly rotating discs that drive the high rotation amplitudes along the major axis while the GCs trace the kinematics of the spheroids of the host galaxies.

For FCC\,083, FCC\,161, FCC\,147, FCC\,167, and FCC\,276, we can compare the red and blue GCs separately. Except for FCC\,147, the red GCs show higher rotation amplitudes and follow the stellar rotation more closely (see also \citealt{Pota2013}). In general, the rotation amplitudes of the blue GCs are smaller. In FCC\,161, FCC\,083, and FCC\,276, the red GCs further have lower velocity dispersions than the blue population.

\begin{figure}
\centering
\includegraphics[width=0.49\textwidth]{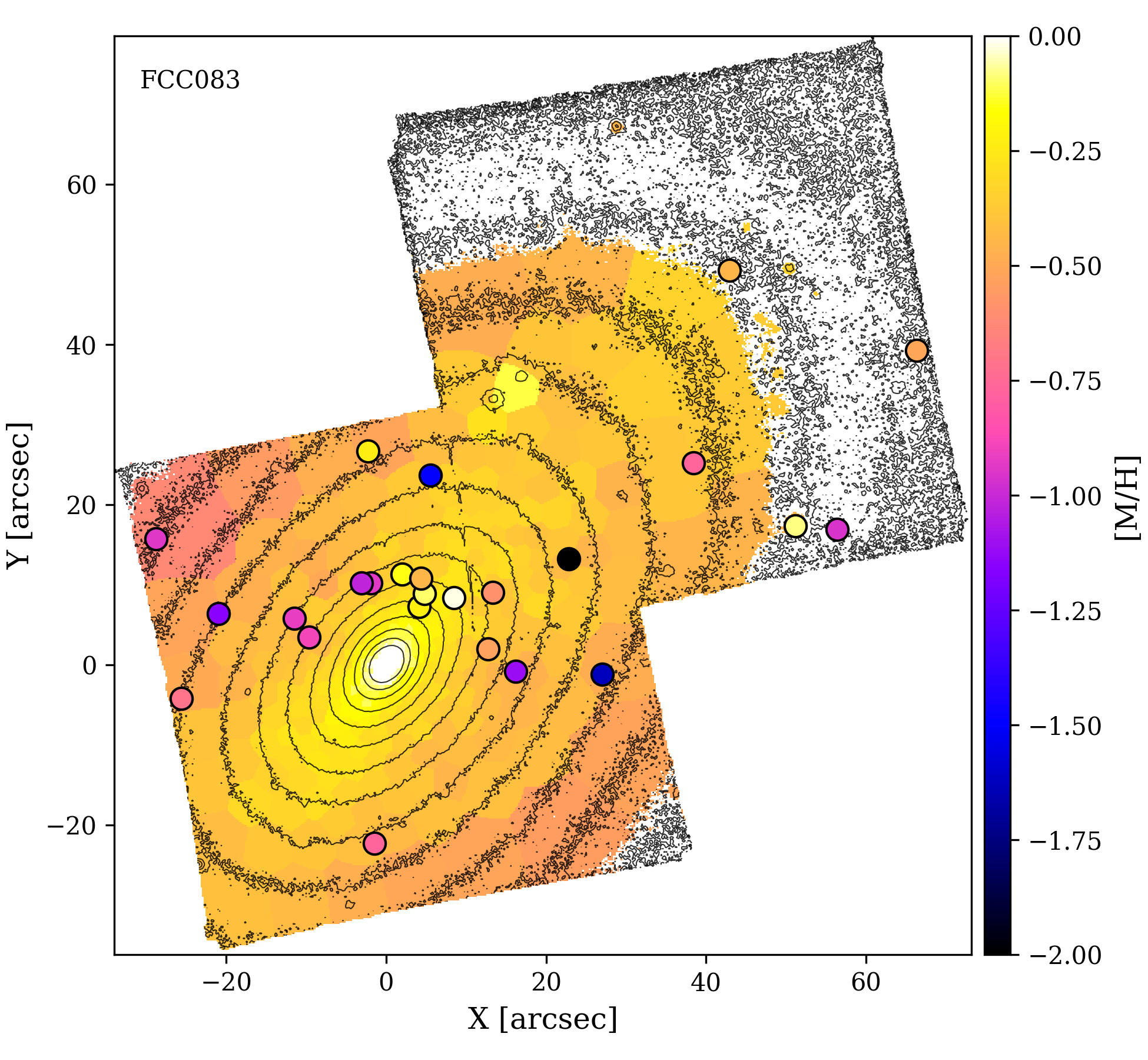}
\includegraphics[width=0.49\textwidth]{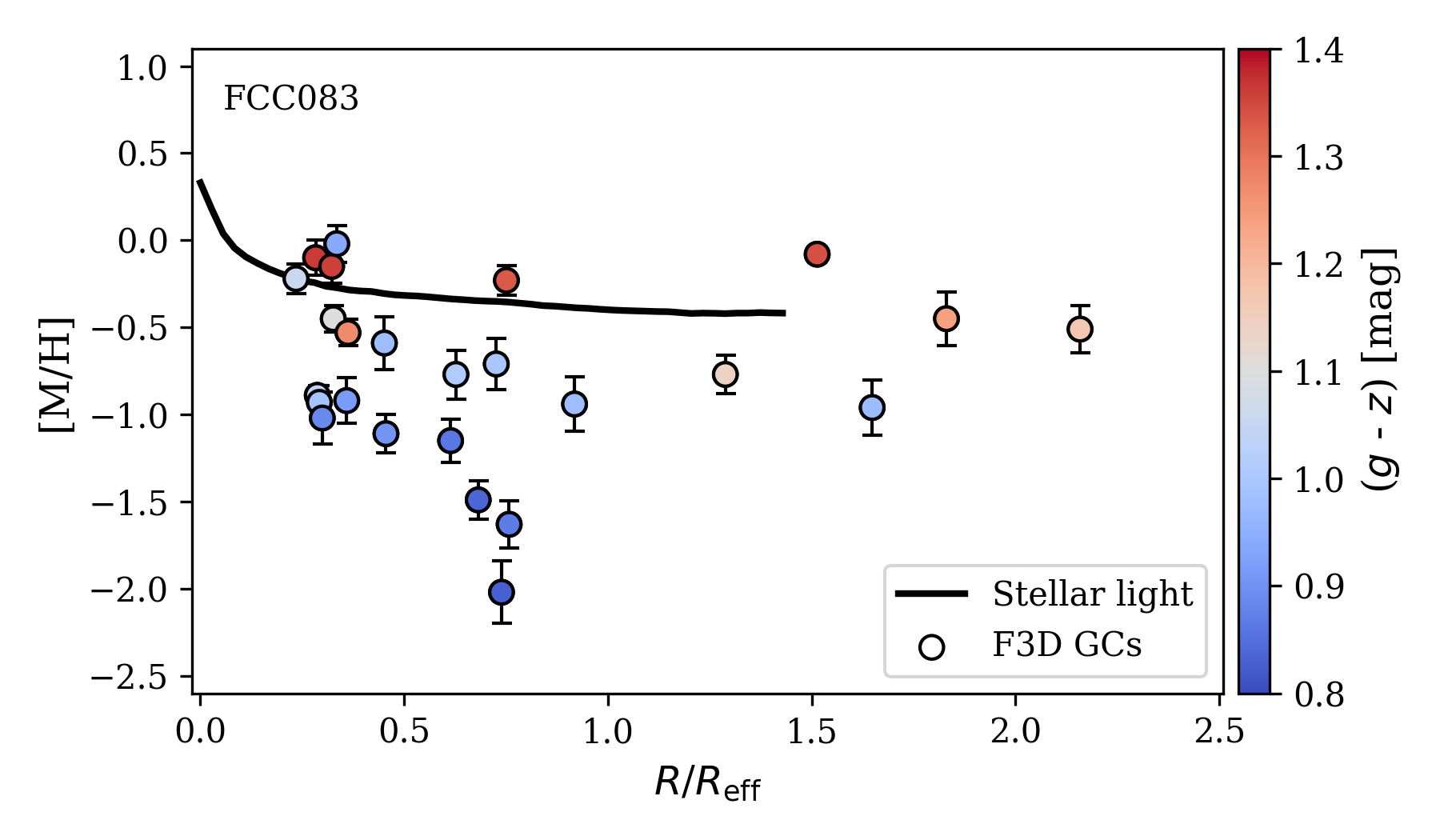}
\caption{Comparison of GC metallicities of FCC\,083 to the host galaxy. \textit{Top}: Metallicity map of FCC\,083 from line-strength measurements with the GCs overplotted overplotted as circles. \textit{Bottom}: radial profile of metallicity with GCs shown as circles, colour-coded by their ($g - z$) colour. The black line shows the radial profile of the metallicity of the galaxy, obtained from the map shown on top.}
\label{fig:metal_example}
\end{figure}

\subsection{Metallicities of globular clusters}
\label{sect:metallicities}

\begin{figure}
\centering
\includegraphics[width=0.49\textwidth]{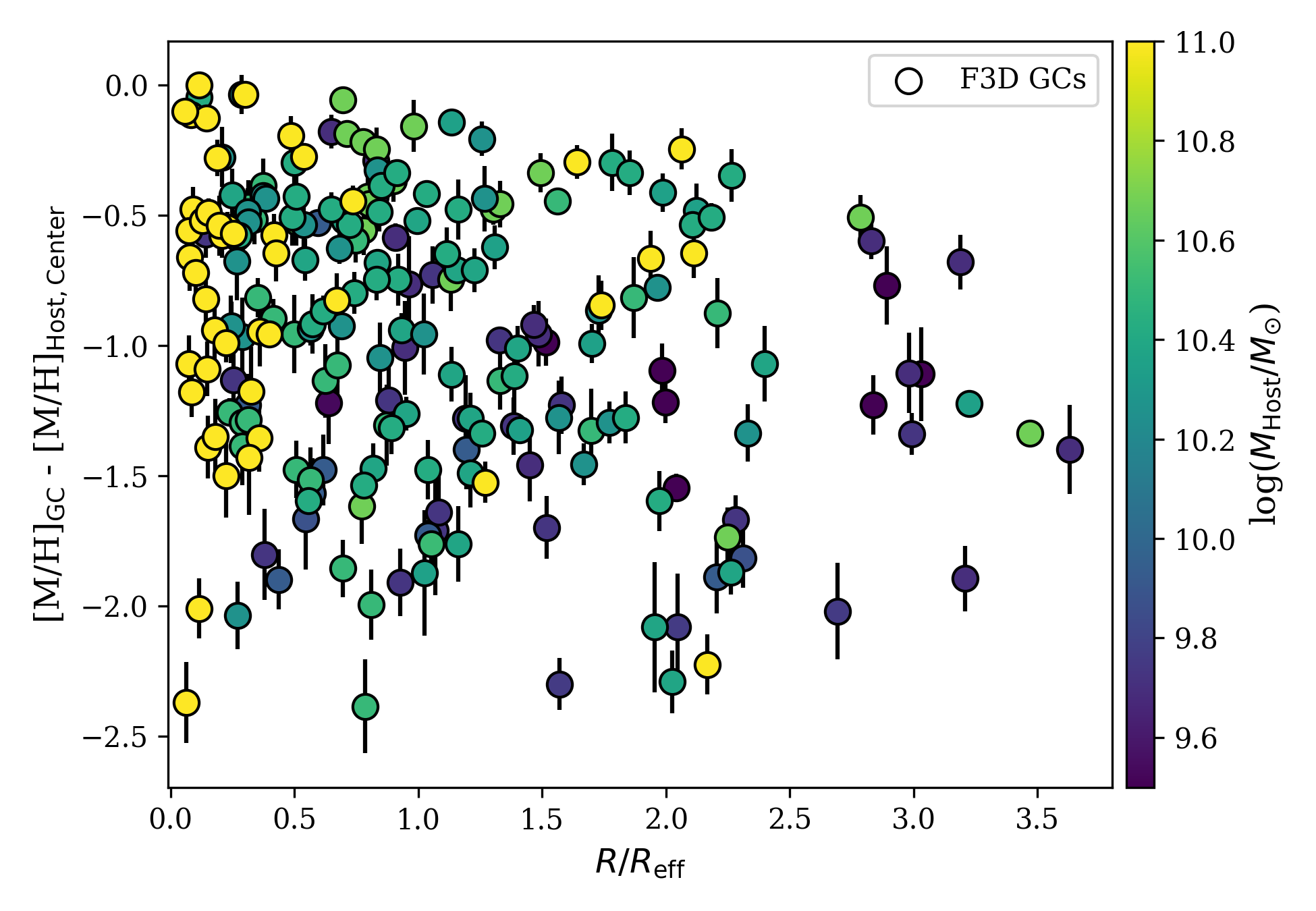}
\caption{Projected radial distribution of F3D GC metallicities from full spectral fitting. GC metallicities relative to the central pixel of the host are shown as a function of GC projected galactocentric distance relative to the host's effective radius for comparison between galaxies. The symbols are colour coded by the host's stellar mass from \cite{Iodice2019} and \cite{Liu2019}. There is one additional GC of FCC\,170 with $R/R_{\text{eff}} \sim 5$ that is not shown in this figure.}
\label{fig:total_met_profile}
\end{figure}

Figure \ref{fig:metal_example} compares the radial distribution of the spectroscopic metallicities of the GCs of FCC\,083 to that of the host galaxy, as an example (see also the right panels in Fig. \ref{fig:maps2} - \ref{fig:maps15} for the remaining galaxies). For this comparison, we used metallicity maps obtained from the line-strength measurements presented in \cite{Iodice2019}.
Since the GC metallicities were obtained from full spectrum fitting, we adjusted possible offsets in the metallicity zero point by fitting the central pixel, but assume a similar gradient. We subtracted this offset from the metallicity maps.

These figures compare the GCs and their host galaxies, but because in some galaxies only a few spectroscopic GC metallicities are available, we combined the sample in Fig. \ref{fig:total_met_profile}. This figure shows a composite radial GC metallicity profile for all F3D galaxies. For better comparison, we show the metallicities relative to the metallicity of the central pixel of the host galaxies and used the projected galactocentric radii relative to the effective radius of the host \citep{Iodice2019, Iodice2019a}. In this figure, the symbols are colour-coded by the stellar mass of the host \citep{Liu2019, Iodice2019}.

\begin{figure*}
\centering
\includegraphics[width=0.99\textwidth]{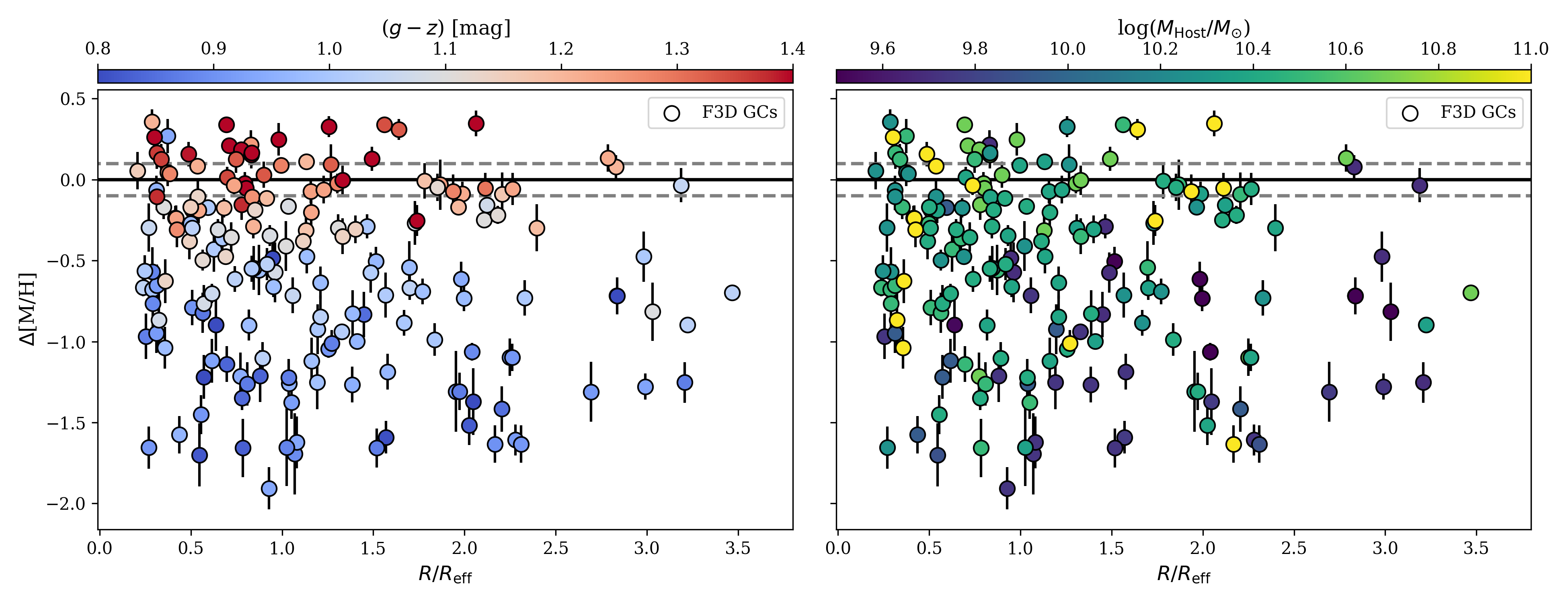}
\caption{Projected radial distribution of the metallicity of the GCs relative to the host metallicity taken from \cite{Iodice2019}, i.e. [M/H]$_\text{GC}$ - [M/H]$_\text{gal}$, colour-coded by the ($g - z$) colour (\textit{left}) and by the host stellar mass (\textit{right}). The black line indicates an offset of zero dex and the dotted lines give the typical scatter found in the host metallicities of $\pm$ 0.1 dex. There are less GCs shown than in Fig. \ref{fig:total_met_profile} because not all F3D galaxies have metallicity maps available.}
\label{fig:tot_met_profile_rel}
\end{figure*}

As Fig. \ref{fig:total_met_profile} shows, the GCs of our sample do not exceed the host's central metallicity, although there are several GCs that appear to be only slightly more metal-poor, even out to one effective radius. While the more massive galaxies appear to have GCs spanning a large range of metallicities and some are even as metal-rich as the centres, the less massive galaxies show GCs that are more metal-poor than their host's centre. This could be connected to the different star formation histories of massive and less massive galaxies. Since massive galaxies might form on shorter time scales than less massive galaxies, they can form metal-rich GCs very early. However, the number of extracted GCs in these low-mass galaxies is quite low and better statistics are needed to address this issue.

While the GCs with the lowest relative metallicities are found over a large range of radii, there seems to be an envelope at high GC metallicities that shows a gradient from $\sim$ 0 to $\sim -0.8$ dex between the centre and 3.5 $R_\text{eff}$ (see Fig. \ref{fig:total_met_profile}). This gradient might be understood as a radial gradient of GC metallicities as it is often observed in individual galaxies using colours as a proxy for metallicity (e.g. \citealt{Harris2016, Caso2017}). However, due to the design of the pointings, the coverage in individual F3D galaxies is not uniform. The GCs of FCC\,213 dominate the inner region in Fig. \ref{fig:total_met_profile} because its four MUSE pointings cover less than 0.4 $R_\text{eff}$, whereas for example the outer pointing of the S0 galaxy FCC\,170 reaches up to 5 $R_\text{eff}$. The mean galactocentric distance of the GCs can be found in Table \ref{tab:F3D_GCs}. In addition, we can only compare projected quantities and therefore, different intrinsic spatial distributions of red and blue GCs can bias the radial profile. It is typically found that the blue GCs are more extended than the red GCs (e.g. \citealt{Harris2016}) and thus it is possible that some of the blue GCs with small projected radii have intrinsic large galactocentric distances.

In addition to comparing the GC metallicities to their host's centre, we can compare them directly to the individual metallicity profiles of the galaxies shown in Fig. \ref{fig:maps1} to \ref{fig:maps15} that were extracted from the metallicity maps. Not every F3D galaxy has a metallicity map available and hence the GCs shown in Fig. \ref{fig:tot_met_profile_rel} are only a sub-sample of those in Fig. \ref{fig:total_met_profile}. We subtracted the host metallicity at each GC position and used the ($g - z$) colour (left panel), and the host stellar mass (right panel) to colour-code the symbols. The dotted line indicates the typical scatter of $\pm$ 0.1 dex in the metallicity profiles of the host.

Fig. \ref{fig:tot_met_profile_rel} illustrates that the red GC population traces the mean stellar metallicity well over a range of radii and host masses, although there is a scatter of $\sim$ 0.5 dex. Over most radii and stellar masses we found GCs that are more metal-rich than the underlying galaxy, but the comparison to Fig. \ref{fig:total_met_profile} shows that those are not more metal-rich than the centres of their host. 
The blue GCs ($g - z < 1.16$ mag) show a broad range of relative metallicites and can be significantly more metal-poor than their hosts at any given radius. The spread in relative metallicities of the blue GCs appears to be independent of the host stellar mass.

\section{Discussion}
\label{sect:discussion}
In the following, we discuss our results on GCs as tracers of kinematics and metallicities. 

\subsection{Globular cluster system rotation}
We built a simple kinematic model for the GCSs in 17 F3D galaxies to derive their rotation amplitude and velocity dispersion. Due to the low numbers of GCs per galaxy, we fixed the systemic velocity of the GC system and assumed that the rotation axis of the GCs coincides with the rotation axis of the host galaxy. Using radial velocities of over 4000 GCs in 27 ETGs of the SLUGGS survey, \cite{Forbes2017} found generally a good agreement between GC mean and host systemic velocity, indicating that the GCs are at rest with respect to their host. For the F3D galaxies with sufficient GC velocities, we also found a good agreement between mean GC and host velocity.

\cite{Pota2013} found that the kinematic position angle of the GCs can differ among red and blue GCs and when compared to the stellar light. We tested the assumption of a common rotation axis with the stars for individual galaxies such as for FCC\,083 or FCC\,161, where a large number of GC velocities are available, and could not find indications for a  kinematic misalignment between GCS and host. However, \cite{Pota2013} found this misalignment at galactocentric distances $>$ 1 $R_e$, where we lack coverage. Nonetheless, the orientation of the MUSE pointings along an axis can bias the results. While most of the pointings where chosen along the major axis, some pointings were placed along the minor axis, for example in FCC\,161 and FCC\,147.
 With a larger MUSE coverage per galaxy, a possible misalignment between GC rotation and the stellar body could be determined, that might indicate a triaxial galaxy shape \citep{Krajnovic2011}.

In our work, we obtained GC velocities only in the inner regions of galaxies. This has the advantage that we can easily compare their velocities to the underlying host, but we are insensitive to possible changes in the GC kinematics at larger radii such as twists in the kinematic position angle. For example, the halo GCs of M31 show a smaller rotation amplitude than the inner GCs \citep{Veljanoski2014}. Also \cite{Pota2013} found variations of ($V/\sigma)_\text{GCS}$ with radius for several of the SLUGGS galaxies. Based on simulations of dry mergers of galaxies, \cite{Bekki2005} predicted an increasing ($V/\sigma$)$_\text{GCS}$ from 2 $R_\text{eff}$ to 6 $R_\text{eff}$, but to detect such signatures, velocity measurements of many outer GCs would be needed. With our simple model, we could only derive radial invariant values for the rotation amplitude and the velocity dispersion and due to the limitations in this model, we refrain from discussing $V/\sigma$ as any bias in the model is boosted in the ratio unnecessarily. We also investigated the angular momentum $\lambda_\text{GCS}$  (e.g. \citealt{Emsellem2007}), but the comparison of it to the stars is complicated due to enhanced uncertainties from $V/\sigma$ and the radial restriction of the GCs.

We found that the velocity dispersion of the GCs closely traces that found in the stars, similar to the findings of \cite{Pota2013} for 12 SLUGGS galaxies. 
This trend is seen also when differentiating between the red and blue GC population, but a separate kinematic analysis of red and blue GCs was not possible in all galaxies. For five galaxies where we could model both the red and blue population separately, we found three galaxies in which the blue GCs have a higher velocity dispersion and low rotation amplitude. This is expected as a result of their accreted origin from random infall directions.

When comparing the rotation amplitudes in the GCSs to the stellar rotation, we found differences between galaxy types. Similar to the results for the SLUGGS ETGs \citep{Pota2013}, we found that the GCs in the elliptical galaxies follow the rotation of the host. In contrast, the GCs of the S0 galaxies in our sample show much lower rotation amplitudes than the stars, especially in the three edge-on S0 galaxies. In these cases, the rotation amplitudes in the stars is maximal due to the inclination angle and is driven by the dynamically cold disc \citep{Pinna2019, Pinna2019b}, whereas the GCs appear to trace the spheroid of the galaxies (e.g. \citealt{Schuberth2010}).

Separating into red and blue GCs, we found that the blue GCs in general show low rotation amplitudes. For the five galaxies, where both populations could be modelled, we found higher rotation amplitudes for the red GCSs in four of them. Typically, it is expected that the red population follows the stellar light closer because of the simultaneous formation of galaxy spheroids and metal-rich GCs. Thus, the metal-rich GCs might be the compact survivors of violent, monolithic-like collapse galaxy formation, also resulting in a higher rotation amplitude \citep{Strader2011}. However, rotation has been detected in both the red and blue GC populations, independently of the host mass or morphology \citep{Arnold2011, Foster2011}. The accreted GC population can show significant rotation if the associated host galaxies fell into the central potential from a preferred direction, as has been discussed in detail for the Local Group \citep{Libeskind2011, Lovell2011}. In simulations, the accreted GC populations are found to show rotation due to a conversion of the orbital angular momentum to intrinsic angular momentum \mbox{\citep{Bekki2005}} and because the angular momentum of the stellar halo is reflected in the angular momentum of the GCs \citep{Veljanoski2016}.

\subsection{Globular cluster metallicities}
We determined spectroscopic metallicity measurements for 238 GCs, a sub-sample of our GC catalogue with high S/N. With this data set, a number of questions can be addressed. For example, in paper II, we will address the colour-metallicity relation of GCs and its implication for galaxy formation in more detail, a topic that is currently discussed in the literature (e.g. \citealt{Usher2012, Harris2017, Villaume2019}). In the following, we focus on the connection between the spectroscopic metallicity of the GC system and of its host galaxy.

We studied how the metallicities of the GCs compare with their host galaxies, both relative to the host's centre and the underlying galaxy at the GC's location. In our limited sample, we found that no GC is more metal-rich than its host's centre. The central metallicity, therefore, not only constitutes the peak of the stellar metallicity as is evident from the radial profiles in Fig. \ref{fig:maps1} - \ref{fig:maps15}, but also of the GC metallicity. Nonetheless, we found several GCs that have metallicities close to their host's central metallicity. These metal-rich GCs are generally located at small projected distances and are associated to massive hosts with $M \gtrsim 10^{10.5} M_\sun$. In the MW, the most metal-rich GCs are found in the bulge. They show a similar metallicity to the bulge stars \citep{Munoz2017, Munoz2018}, indicating an in-situ formation together with the bulge.

In our sample of F3D GC metallicities, we found GCs in the range between [M/H] $\sim -2.3$ dex to $\sim +0.4$ dex, fully exploiting the metallicity grid of the SSP models. We do not expect to miss GCs at lower metallicities due to the empirical metallicity floor of GCs at $\sim -2.5$ dex (e.g. \citealt{Beasley2019}).
The more massive galaxies of our sample ($M \gtrsim10^{10} M_\sun$) show the largest spread of GC metallicities from high metallicities comparable to the host's centre down to the very metal-poor regime, whereas the less massive F3D galaxies show a smaller spread of GC metallicities and seem to have only GCs that are more metal-poor than their host's centre. Although this difference might not be significant due to low numbers of GCs in the low mass galaxies, this is consistent with the finding that the relative number of red GCs decreases with galaxy mass and thus low mass galaxies only have blue GCs \citep{Peng2006}.
However, the number of GCs in low mass F3D galaxies are low, with only 1 -- 3 GCs per galaxy. 

Broad GC metallicity ranges have been found in many massive galaxies, for example in M31 (e.g. \citealt{Barmby2000, Perrett2002, Caldwell2016}), Centaurus A \citep{Woodley2010b}, Sombrero galaxy \citep{AlvesBrito2011}, and M87 \citep{Strader2011, Villaume2019}. The broad range of GC metallicities is often connected to the evolutionary history of the host galaxy with the red, most-metal rich GCs having formed in-situ and blue GCs having been accreted from more metal-poor dwarf galaxies during the assembly of the host. From this scenario naturally follows that more massive galaxies with richer merger histories also acquire a broad GC metallicity distribution due to the accretion of satellites of different masses (e.g. \citealt{Kravtsov2005, Tonini2013, Li2014, Choksi2018, Kruijssen2019, Kruijssen2019b}). Conversely, low-mass galaxies that are thought to have only a limited number of mergers in their past, then obtain a narrower GC metallicity distribution. 

Comparing the GC metallicities to the underlying host metallicity profile has shown that there are several red GCs that are more metal-rich than their host locally. Especially FCC\,184, a lenticular galaxy with a prominent bar in the centre, shows a large number of these GCs. At the same time, FCC\,184 stood out as an outlier with its red GCs having a low velocity dispersion that is also lower than that of the stars. Together with the large number of very metal-rich GCs, this might indicate that these metal-rich GCs were formed in a disc that is viewed face-on. However, more rigorous modelling of the GC kinematics and stellar populations would be required to test this.

Also other galaxies have GCs that are more metal-rich then the host locally. Besides projection effects, this could be explained by different star and cluster formation conditions, if the GCs were born closer to the host centre and were ejected to larger distances \citep{Leung2019}, or as a result of violent interactions during the initial collapse and the formation from clumps (e.g. \citealt{Kruijssen2015}).
In addition, their existence could indicate a more metal-rich population within the galaxy that is not evident from the mean metallicity profile of the host \mbox{\citep{Fahrion2019b}}. In a hierarchical assembly history, it is possible that such a metal-rich component including GCs formed from fast self-enrichment while the more metal-poor component was acquired during the accretion of metal-poor dwarf galaxies. 

Besides an early formation in the parent halo, it has been suggested that metal-rich GCs form during major mergers of gas rich galaxies (e.g \citealt{Li2014, Choski2018}), which could explained their extended radial distribution. Moreover, some of the GCs that are more metal-rich then their host locally might be ultra compact dwarf galaxies (UCDs) rather than classical GCs. UCDs are often discussed to be the remnant nuclear star clusters of disrupted galaxies (e.g. \citealt{Bekki2003, Drinkwater2003, Pfeffer2013, Strader2013, Fahrion2019}), and because those can be more metal-rich then their host galaxy and the surrounding GC system (e.g. \citealt{Paudel2011}), finding a GC more metal-rich than the host at large separations could indicate such a remnant nuclear star cluster.

Red GCs are often used as tracers of the in-situ halo metallicity. For example, \cite{Beasley2008} found a good agreement of the metallicity distribution function of the red GCs of Centaurus\,A with that of its halo stars, while the studied GCs appear to be more metal-poor than the inner regions of the galaxy. A similar observation was made for the brightest cluster galaxy NGC\,6166, where the metal-richer GCs are found to closely follow the radial distribution, ellipticity and mean metallicity of the halo light \citep{Harris2016}. Using hydrodynamical simulations, \cite{ForbesRemus2018} studied the metallicity gradients of in-situ and accreted GCs, and although they appear similar to that of the metal-rich and metal-poor GCs, they found that a one-to-one connection between metal-rich and in-situ formation is not given because major mergers bring in both metal-rich and metal-poor GCs. Depending on the mass of the satellite, the accreted GCs are deposited in different regions in the galaxy. Massive satellites can therefore deposit their metal-rich GCs in the inner regions, while the generally more metal-poor GCs of low mass satellites end up at larger galactocentric distances.

With F3D, we have the opportunity to extend the comparison to the inner parts of galaxies ($< 1\,R_\text{eff}$). Using the metallicities from line strength analysis as presented in \cite{Iodice2019}, we investigated the relative metallicity between GCs and host. The red GC population was found to trace the host metallicity closely over all studied radii and stellar host masses, indicating a co-evolution of metal-rich GCs and the host galaxy. The blue GCs, however, appeared to be more metal-poor than the host, even at large projected distances. Similar observations were made when comparing halo stars and halo GCs directly as described by \cite{Lamers2017}, for example in the MW \citep{Ryan1991, An2012, Harris2016b}, Centaurus\,A \citep{Rejkuba2005, Rejkuba2014, Beasley2008, Crnojevic2013} or even the Fornax dwarf spheroidal galaxy \citep{Larsen2012}.
\cite{Lamers2017} argued that the metallicity contrast between GCs and halo stars is a result of galaxy assembly via mergers during which more metal-poor GCs are more likely to survive until today because of their origin in metal-poor dwarfs, where the destruction via shocks is less likely. Also, accreted GCs represent the formation conditions in their parent galaxy before the time of accretion.

\section{Conclusions}
\label{sect:conclusions}
In this work, we present a catalogue of 722 spectroscopically confirmed GCs of 32 galaxies in the Fornax cluster. We give the LOS velocities for these GCs and the metallicity for 238 GCs.
We summarise our results as follows.
\begin{itemize}
\item{By subtracting a MGE model of the galaxies from their collapsed MUSE images, we detected GCs in the inner regions of galaxies. After cross-referencing with the ACSFCS catalogue of GC candidates based on photometry and sizes \citep{Jordan2015}, we extracted the spectra of the GCs using a PSF-weighted circular aperture and subtracted the spectrum of the underlying galaxy.}

\item{We classified the GCs based on their spectral S/N for further analysis. The S/N depends on a combination of the intrinsic brightness of the GC, the exposure time of the observation, the contrast with the underlying galaxy and the metallicity. We determined the LOS velocities for GCs with \mbox{S/N $\geq$ 3 \AA$^{-1}$} and also metallicities were obtained from full spectral fitting for the GCs with \mbox{S/N $\geq$ 8 \AA$^{-1}$}. Depending on the S/N, the resulting random uncertainties are in the range of \mbox{5 $\lesssim \delta \text{v} \lesssim 60$ km s$^{-1}$} and \mbox{0.05 $\lesssim \delta \text{[M/H]} \lesssim 0.20$ dex}, respectively.}

\item{Using the GC LOS velocities, we modelled the rotation amplitude and velocity dispersion of the GC systems in several galaxies. Where possible, we modelled the red and blue GCs separately. We found that the GC velocity dispersion traces that of the stars. For the elliptical galaxies in our sample, especially the red GCs follow the stellar rotation while the GC rotation is lower than in the stars for the lenticular galaxies. This illustrates that the GCs follow the kinematics of the spheroid of the host galaxy rather than the disc. The blue GCs generally show low rotation amplitudes.}

\item{We compared the GC metallicities to the centres of the respective host galaxies and found that the central metallicity not only sets an upper limit to the stellar metallicity, but also to the GC metallicity. Although no GC exceeds the metallicity set by their host's centre, we found several with similar metallicities, even at distances $\sim 1\,R_\text{eff}$. The more massive galaxies in our sample show a large spread in metallicities, from very metal-poor ($\lesssim -2.3$ dex) to super-solar values, while the less massive galaxies show a narrower distribution in GC metallicities that are on average more metal-poor.}

\item{Comparing the GC metallicities to the metallicity of the stellar body locally showed that the red GCs generally trace the stellar metallicity profile closely from the inner regions out to a few $R_\text{eff}$. The blue GCs, however, are more metal-poor than the host galaxies, even in the outer regions. This was found independent of the galaxy stellar mass.}
\end{itemize}

In accordance with other studies, we found that GCs are valuable tracers of the enclosed mass of their host galaxy. Especially the red GCs further trace the kinematics of the galaxy spheroid and follow the galaxy metallicity from the inner parts out into the halo regions. Contrary, the blue, metal-poor GCs show larger deviations with respect to the properties of the host, independent of the host stellar mass. This method of extracting high quality GC spectra provides a new efficient way to assess accurate membership and chemodynamical properties of GC systems (especially in the inner regions), which will be crucial to exploit the wealth of IFU data of galaxies in the local Universe.

\begin{acknowledgements}
We thank the anonymous referee for helpful comments and suggestions that have improved this manuscript. KF thanks Eric Emsellem for helpful discussions. GvdV acknowledges funding from the European Research Council (ERC) under the European Union's Horizon 2020 research and innovation programme under grant agreement No 724857 (Consolidator Grant ArcheoDyn). J. F-B  acknowledges support through the RAVET project by the grant AYA2016-77237-C3-1- P from the Spanish Ministry of Science, Innovation and Universities (MCIU) and through the IAC project TRACES which is partially supported through the state budget and the regional budget of the Consejer\'{i}a de Econom\'{i}a, Industria, Comercio y Conocimiento of the Canary Islands Autonomous Community. IMN acknowledges support from the AYA2016-77237-C3-1-P grant from the Spanish Ministry of Economy and Competitiveness (MINECO) and from the Marie Sk\l odowska-Curie Individual {\it SPanD} Fellowship 702607. EMC is supported by MIUR grant PRIN 2017 20173ML3WW\_001 and by Padua University grants DOR1715817/17, DOR1885254/18, and DOR1935272/19. This research made use of Photutils, an Astropy package for detection and photometry of astronomical sources \citep{Bradley2019}. This research made use of Astropy,\footnote{\url{http://www.astropy.org}} a community-developed core Python package for Astronomy \citep{astropy:2013, astropy:2018}.
\end{acknowledgements}

\bibliographystyle{aa} 
\bibliography{References}

\appendix
\section{Catalogue of globular clusters}
\label{app:catalogue}
Table \ref{tab:F3D_GCs_cat} gives an overview of the catalogue of GCs that will be available online in its full form. For every GC, we give the host galaxy, the coordinates, the $g$ and $z$ magnitudes from the ACSFCS \citep{Jordan2015} if available, the synthetic MUSE ($g - z$) colour, the spectral S/N at 6000 \AA, the LOS velocity with random uncertainty, the metallicity, and its uncertainty. The synthetic MUSE colours were determined by applying the HST ACS F450W ($g$) and F850LP ($z$) transmission curves to the MUSE spectra. For high S/N, this colour estimate is quite reliable, but at lower S/N, large deviations between ($g - z$)$_\text{ACSFCS}$ and $(g - z)_\text{MUSE}$ can occur, possibly due to the presence of sky residual and telluric lines in the MUSE spectrum. Also, the MUSE wavelength range does not cover the F850LP filter completely.

\begin{table*}
\centering
\caption{Excerpt of the F3D GC catalogue for FCC\,083. The complete catalogue will be available online.}
\label{tab:F3D_GCs_cat}
\begin{tabular}{c c c c c c c c c c c}\hline
Host & RA & DEC & $z_\text{ACSFCS}$ & $g_\text{ACSFCS}$ & $(g - z)_\text{MUSE}$ & S/N & $v_\text{LOS}$  & $\delta v_\text{LOS}$ & [M/H] & $\delta$[M/H] \\
 & (J2000) & (J2000) & (mag) & (mag) & (mag) & (\AA$^{-1}$) & (km s$^{-1}$) & (km s$^{-1}$) & (dex) & (dex) \\ 
 (1) & (2) & (3) & (4) & (5) & (6) & (7) & (8) & (9) & (10) & (11) \\
 \hline\hline
FCC083 & 52.644913 & $-$34.850722 & 22.13 & 23.49 & 1.13 & 9.7 & 1643.45 & 7.12 & $-$0.15 & 0.10 \\
FCC083 & 52.644133 & $-$34.850864 & 20.29 & 21.39 & 1.08 & 22.7 & 1713.48 & 4.85 & $-$0.45 & 0.08 \\
FCC083 & 52.642746 & $-$34.851560 & 22.12 & 23.06 & 1.15 & 11.6 & 1669.65 & 11.4 & $-$0.02 & 0.11 \\
FCC083 & 52.641295 & $-$34.853339 & 21.22 & 22.49 & 1.20 & 9.2 & 1708.91 & 6.28 & $-$0.53 & 0.08 \\
FCC083 & 52.649481 & $-$34.852272 & 21.42 & 22.34 & 0.84 & 10.1 & 1722.16 & 10.12 & $-$0.92 & 0.13 \\
FCC083 & 52.641143 & $-$34.855731 & 21.54 & 22.56 & 0.99 & 6.8 & 1547.31 & 9.47 & -- & -- \\
FCC083 & 52.640567 & $-$34.855252 & 22.34 & 23.24 & 0.87 & 4.8 & 1635.36 & 22.96 & -- & -- \\
FCC083 & 52.641085 & $-$34.851349 & 21.65 & 22.63 & 1.04 & 9.6 & 1543.29 & 7.74 & $-$0.59 & 0.15 \\
FCC083 & 52.640120 & $-$34.854123 & 21.06 & 21.97 & 0.93 & 14.5 & 1595.18 & 6.2 & $-$1.11 & 0.11 \\
\hline
\end{tabular}
\tablefoot{(1) Galaxy name from \cite{Ferguson1989}. (2) and (3): Right ascension and declination. (4), (5): $g$ and $z$-band magnitudes from \cite{Jordan2015} and (6) synthetic ($g - z$) colour from the MUSE spectrum. (7) spectral S/N. (8), (9) GC LOS velocity and random uncertainty, (10), (11) GC metallicity and random uncertainty.
}
\end{table*}

\section{Globular cluster velocities and metallicities in comparison to their host galaxies}
\label{app:all_the_plots}
\begin{figure*}

\includegraphics[width=0.45\textwidth]{FCC083_GCs_vel_img.png}
\includegraphics[width=0.45\textwidth]{FCC083_GCs_met_img.png}\\
\includegraphics[width=0.45\textwidth]{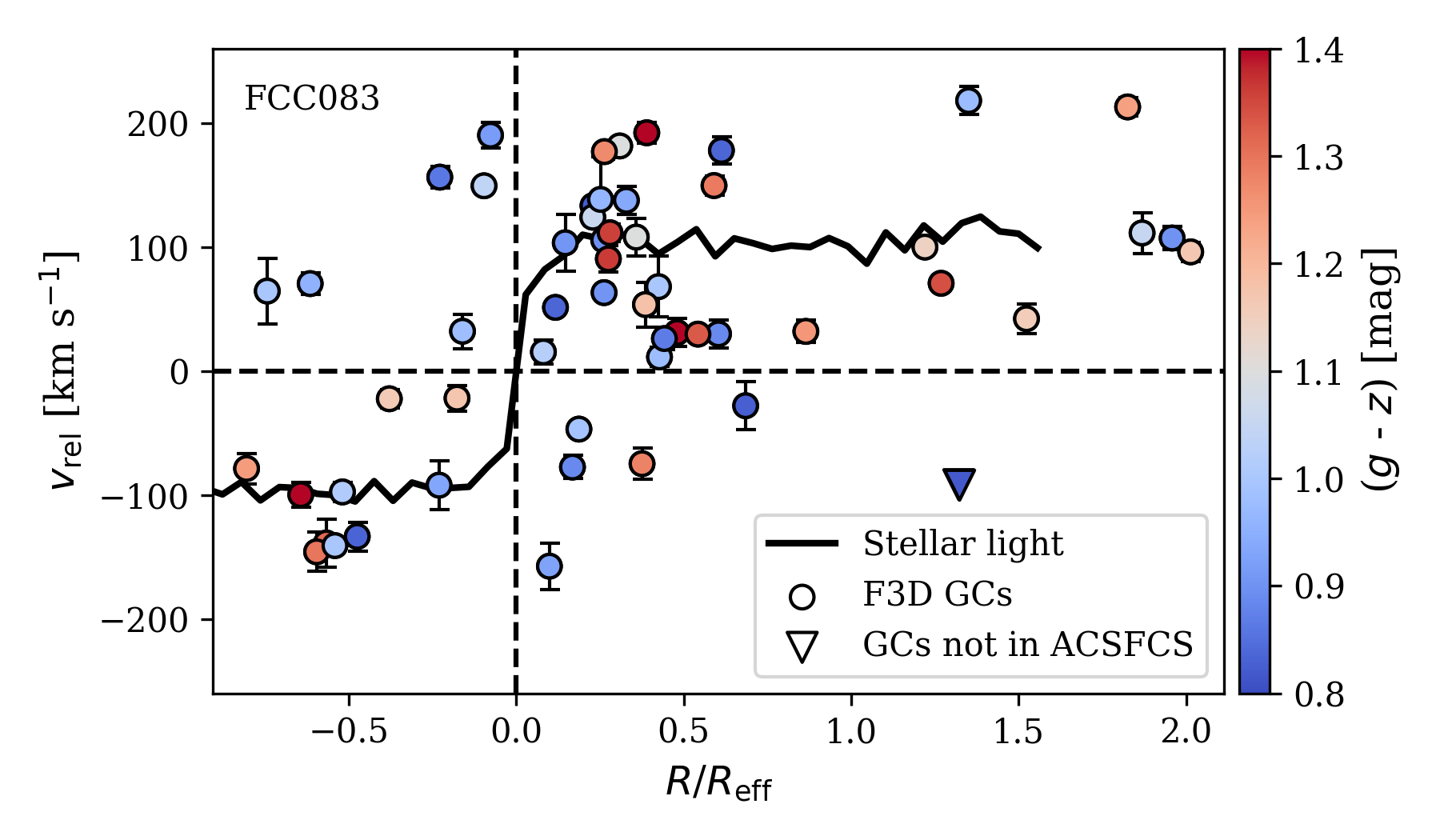}
\includegraphics[width=0.45\textwidth]{FCC083_GC_met_profile.png}
\caption{GCs of FCC\,083 in comparison to the host galaxy. \textit{Top panels}: GC velocities (\textit{left}) and metallicites (\textit{right}) shown as circles on top of the Voronoi-binned maps presented in \citep{Iodice2019}. The black contours indicate the surface brightness levels from the original MUSE cube to guide the eye. \textit{Bottom left}: GC LOS velocities versus galactocentric distance projected on the major axis, and \textit{bottom right}: radial GC metallicity profile. The circles show the GCs, colour-coded by ($g - z$) from the ACSFCS. GCs that were not covered in the catalogue of  \cite{Jordan2015} are shown by triangles. The black line refers to the profile from the stellar light.}
\label{fig:maps1}
\end{figure*}

We show the F3D GCs in comparison to their hosts for every galaxy in Fig. \ref{fig:maps1} to \ref{fig:maps15}. In the left panels of these figures, the GC LOS velocities are shown on top of the LOS velocity maps from \cite{Iodice2019}. Because different fitting procedures can result in small systematic offsets of the LOS velocity, we adjusted possible offsets by separately fitting the central pixel of each map with the same \textsc{pPXF} setup that was used for the GCs. In addition to the maps, we also show the radial profile of host and GC velocities by projecting their position onto the kinematic axis of the galaxy. The rotation curve is also obtained along the kinematic major axis.

The right panels show a similar comparison for the metallicities of the GCs and the host. For the host galaxies, we used metallicities inferred from line-strength measurements \citep{Iodice2019}. Again, we corrected for possible offsets in the metallicities due to different measurement techniques by fitting the central pixel. The bottom panels show the radial metallicity profile for each galaxy.

\textit{Due to size restrictions, we only show a few examples here. The full appendix is available upon request or in the journal article.}

\FloatBarrier

\begin{figure*}

\includegraphics[width=0.45\textwidth]{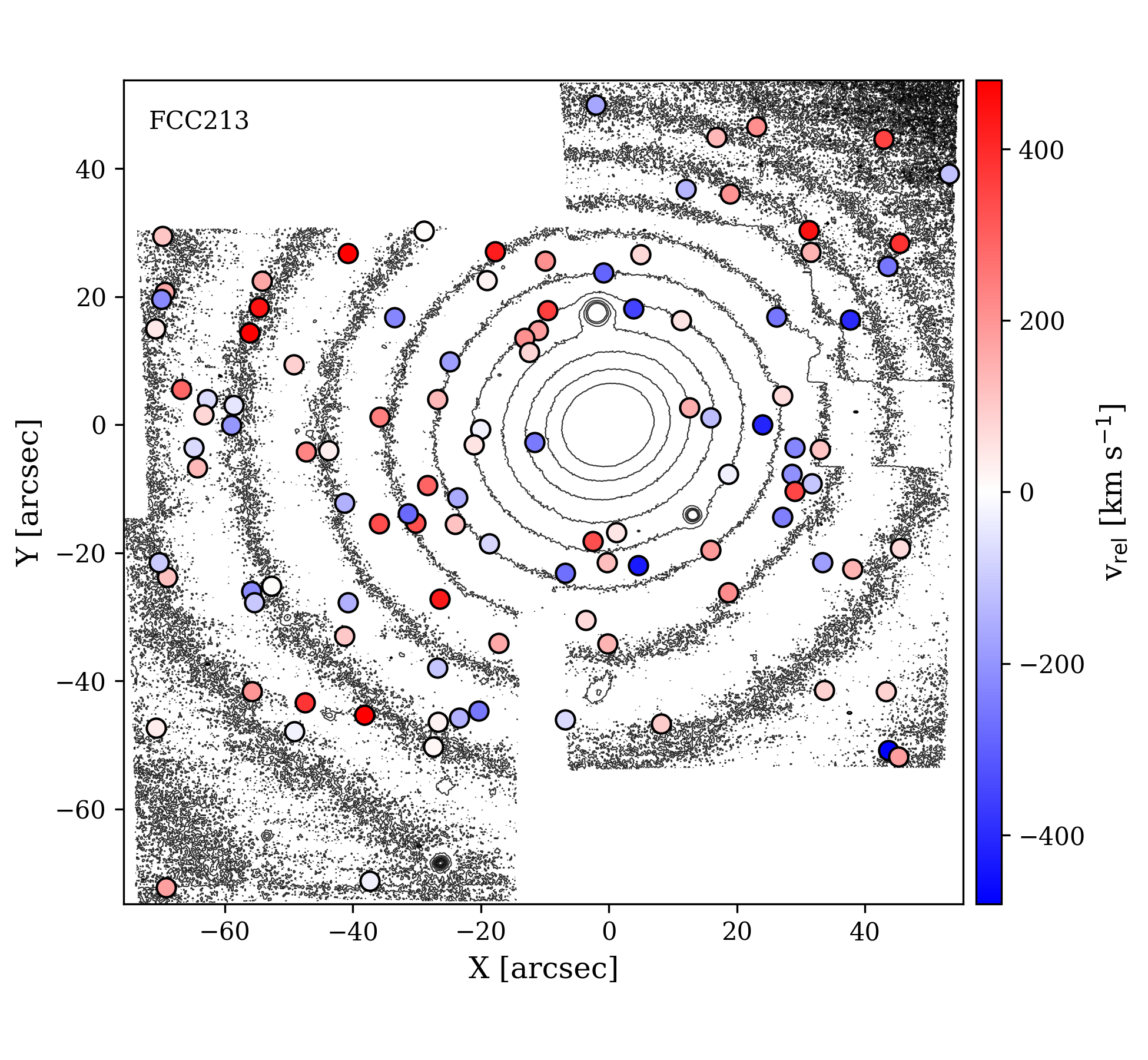}
\includegraphics[width=0.45\textwidth]{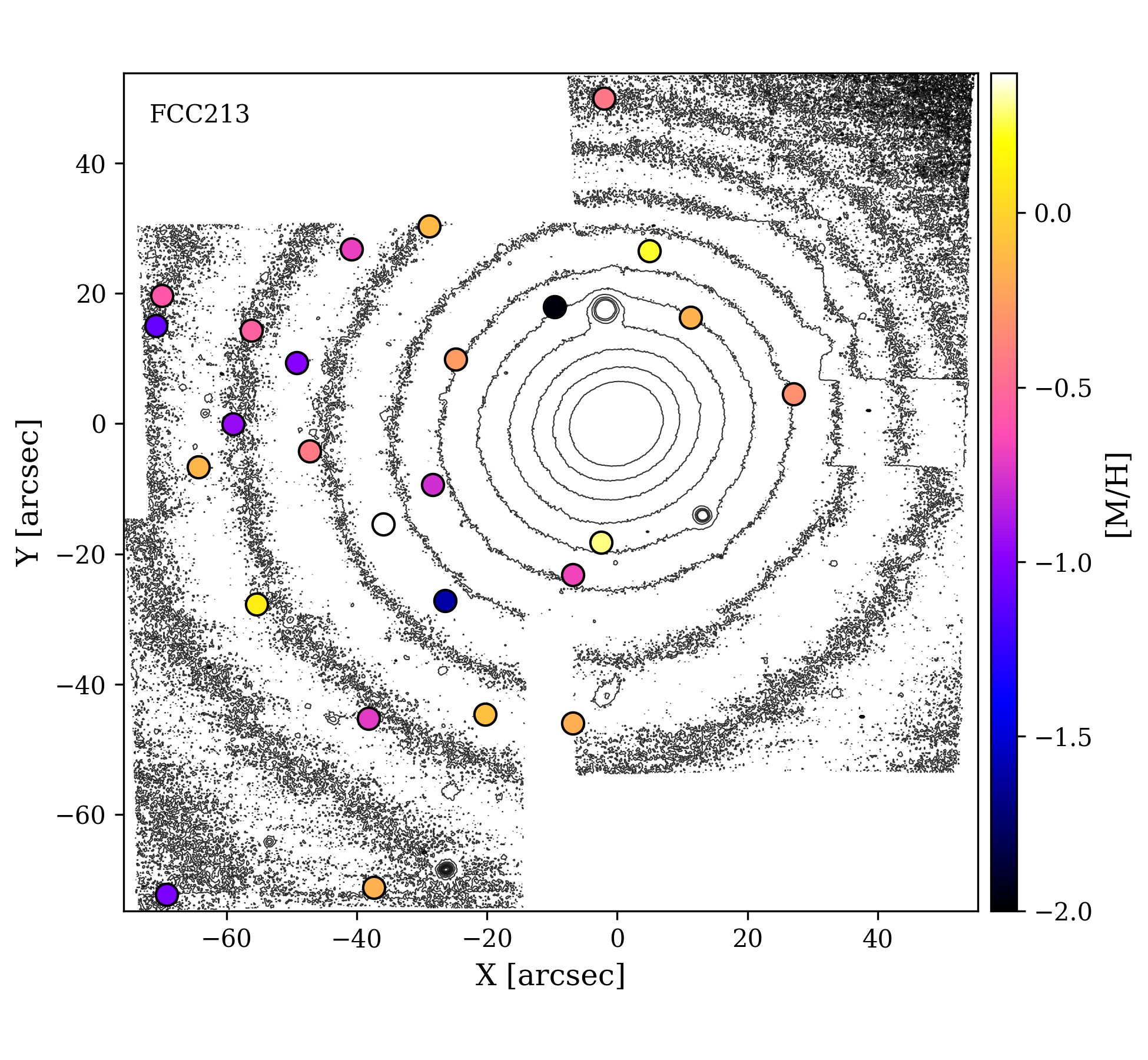}\\
\includegraphics[width=0.45\textwidth]{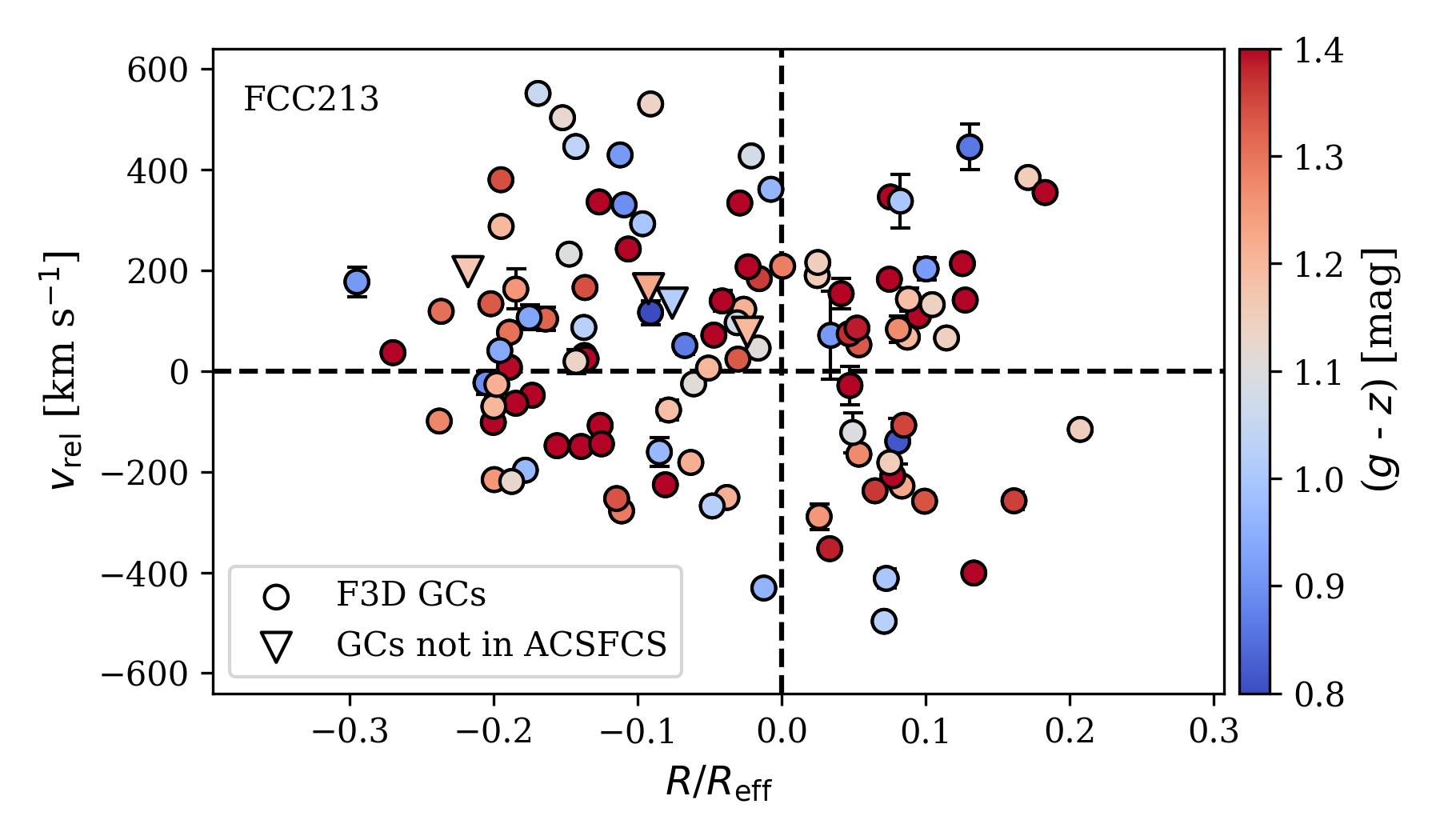}
\includegraphics[width=0.45\textwidth]{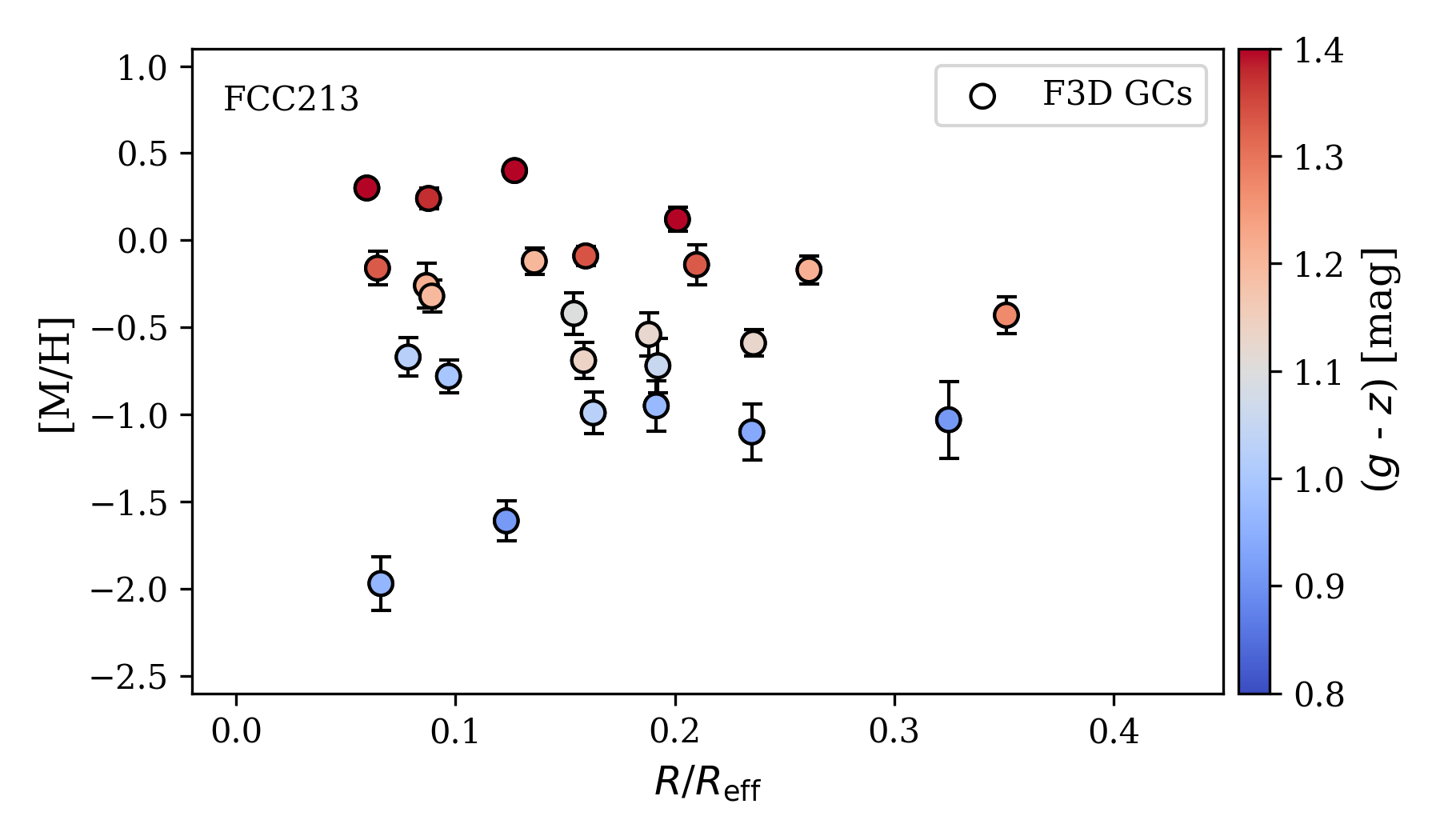}\\

\includegraphics[width=0.45\textwidth]{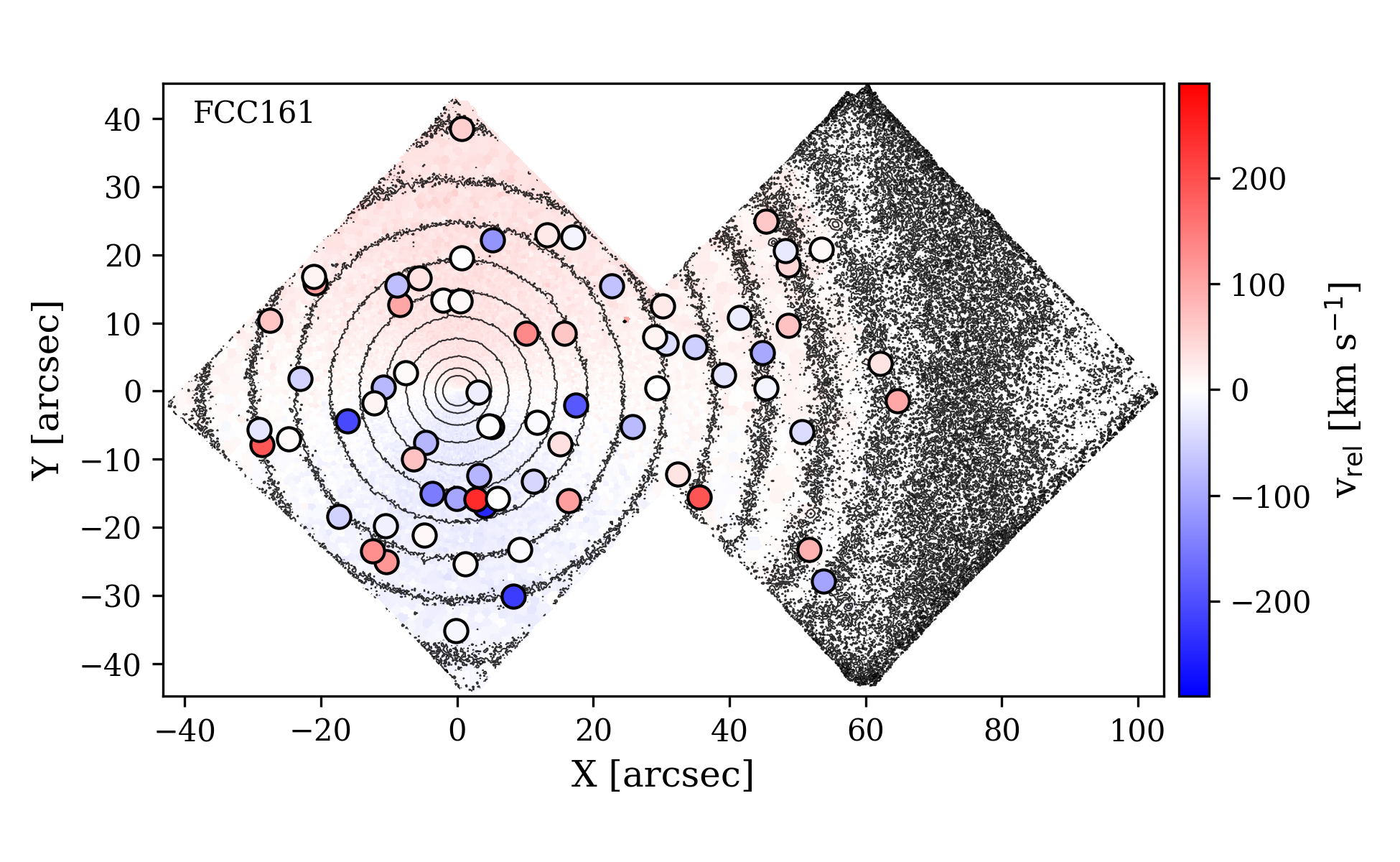}
\includegraphics[width=0.45\textwidth]{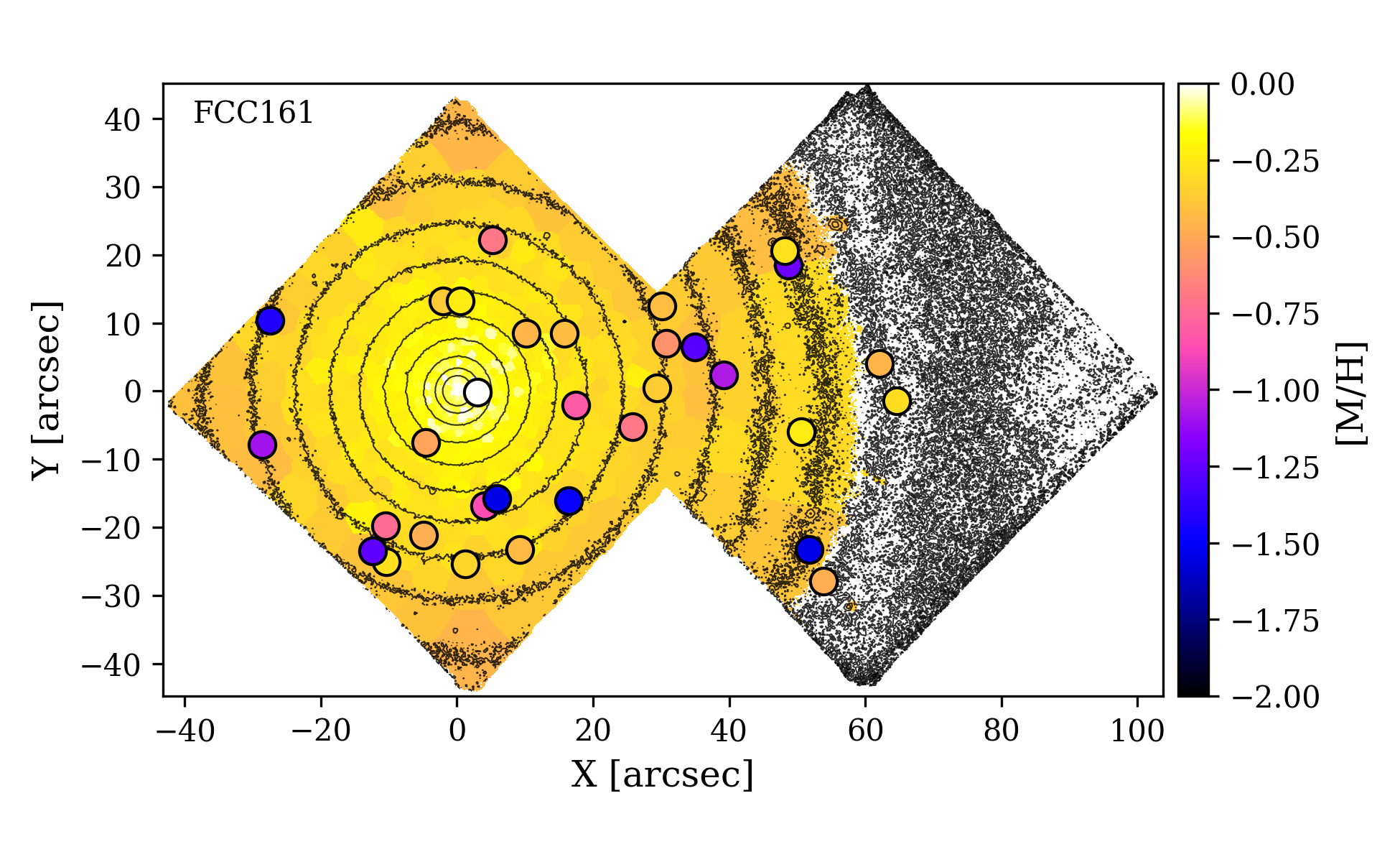}\\
\includegraphics[width=0.45\textwidth]{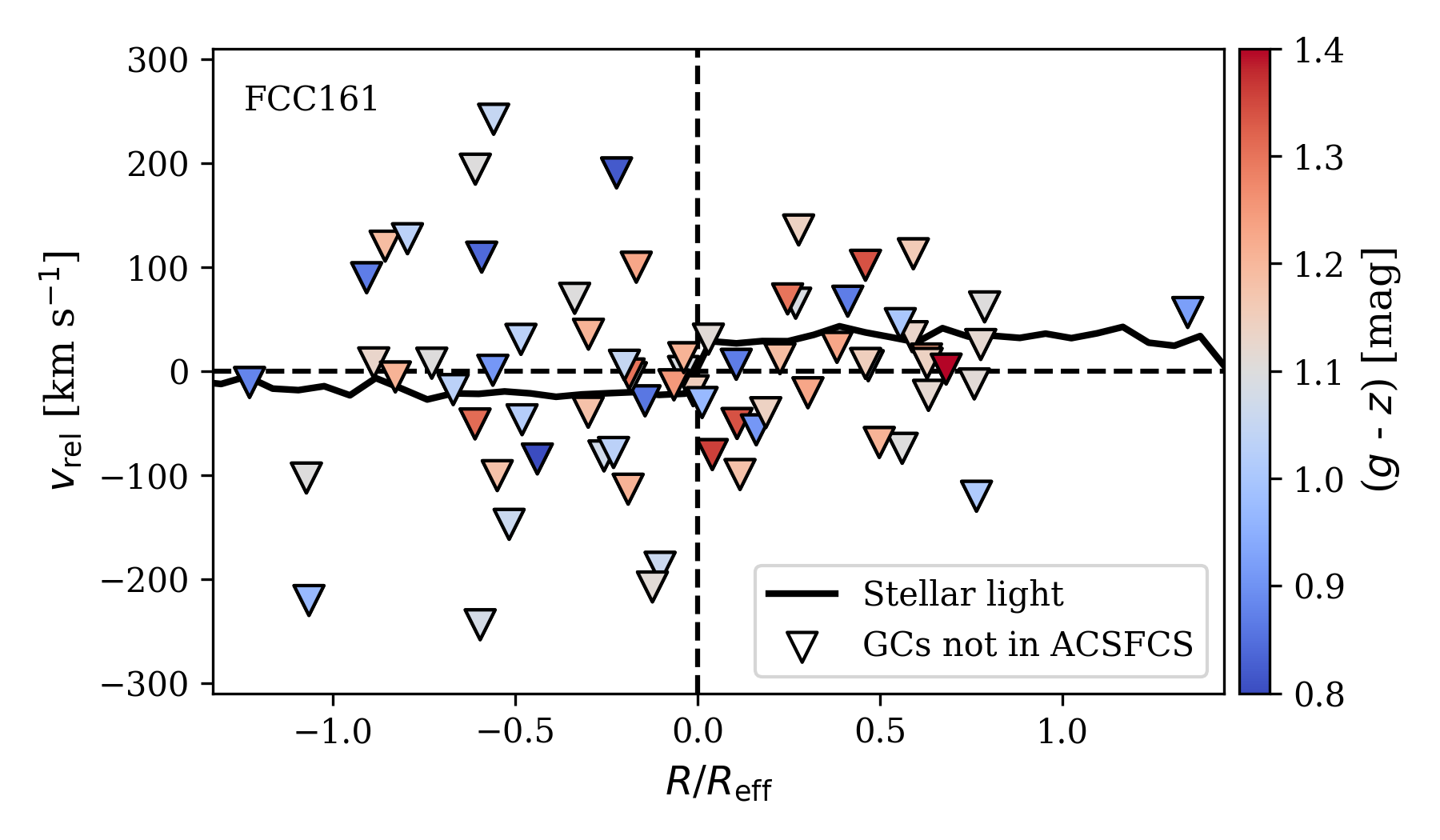}
\includegraphics[width=0.45\textwidth]{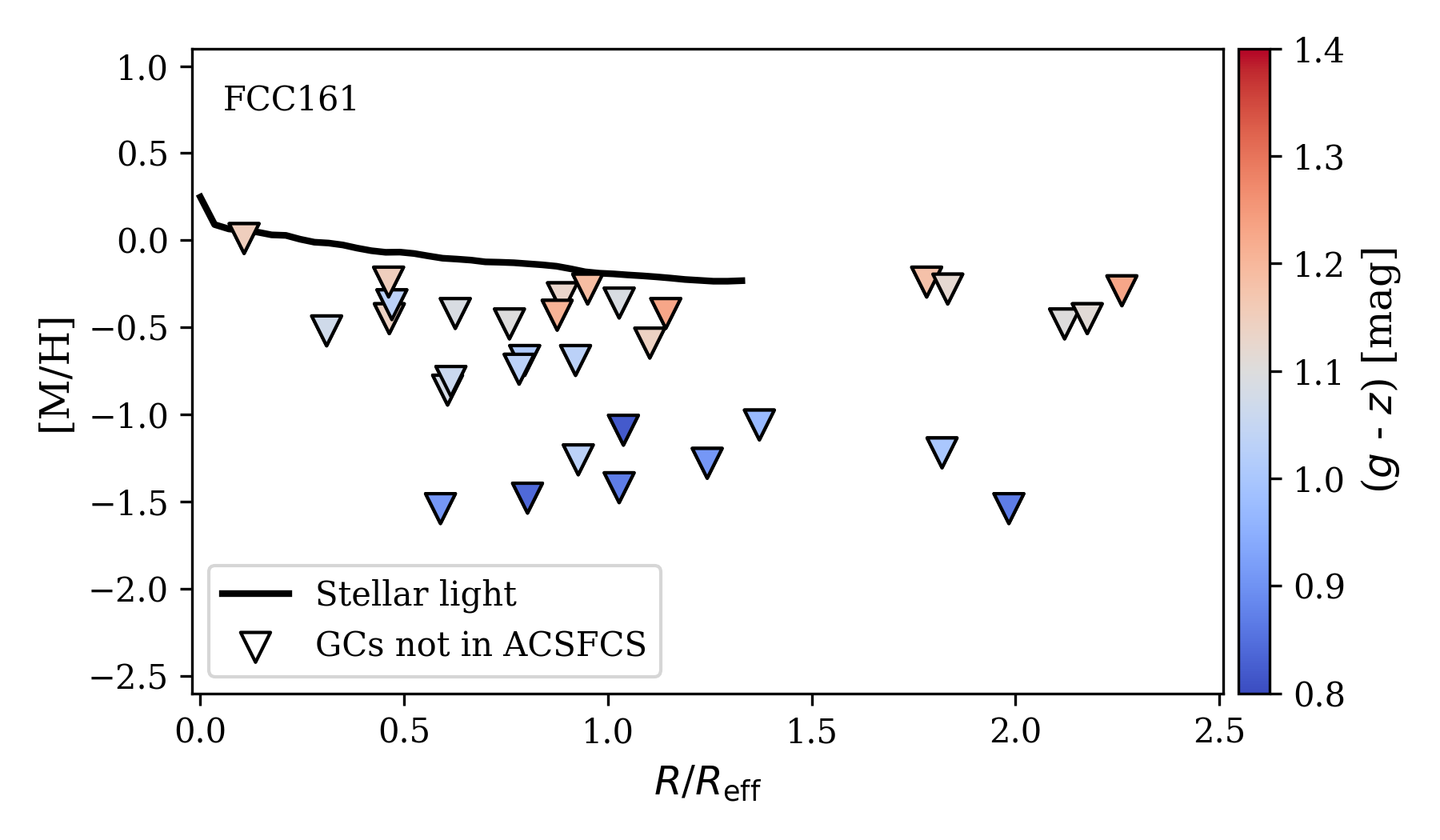}\\
\caption{Same as Fig. \ref{fig:maps1}, but for FCC\,213 (NGC\,1399 and FCC\,161.}
\label{fig:maps5}
\end{figure*}

\begin{figure*}

\includegraphics[width=0.45\textwidth]{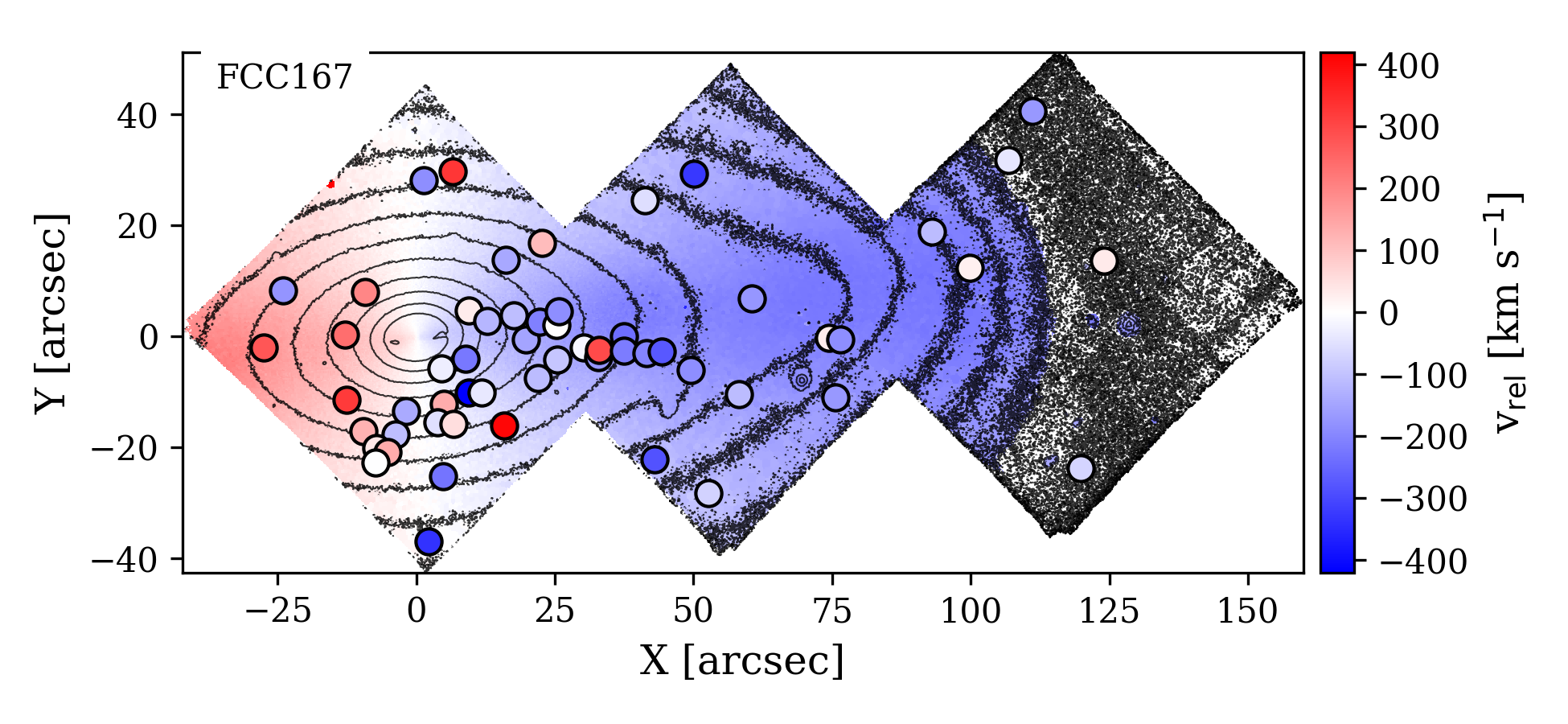}
\includegraphics[width=0.45\textwidth]{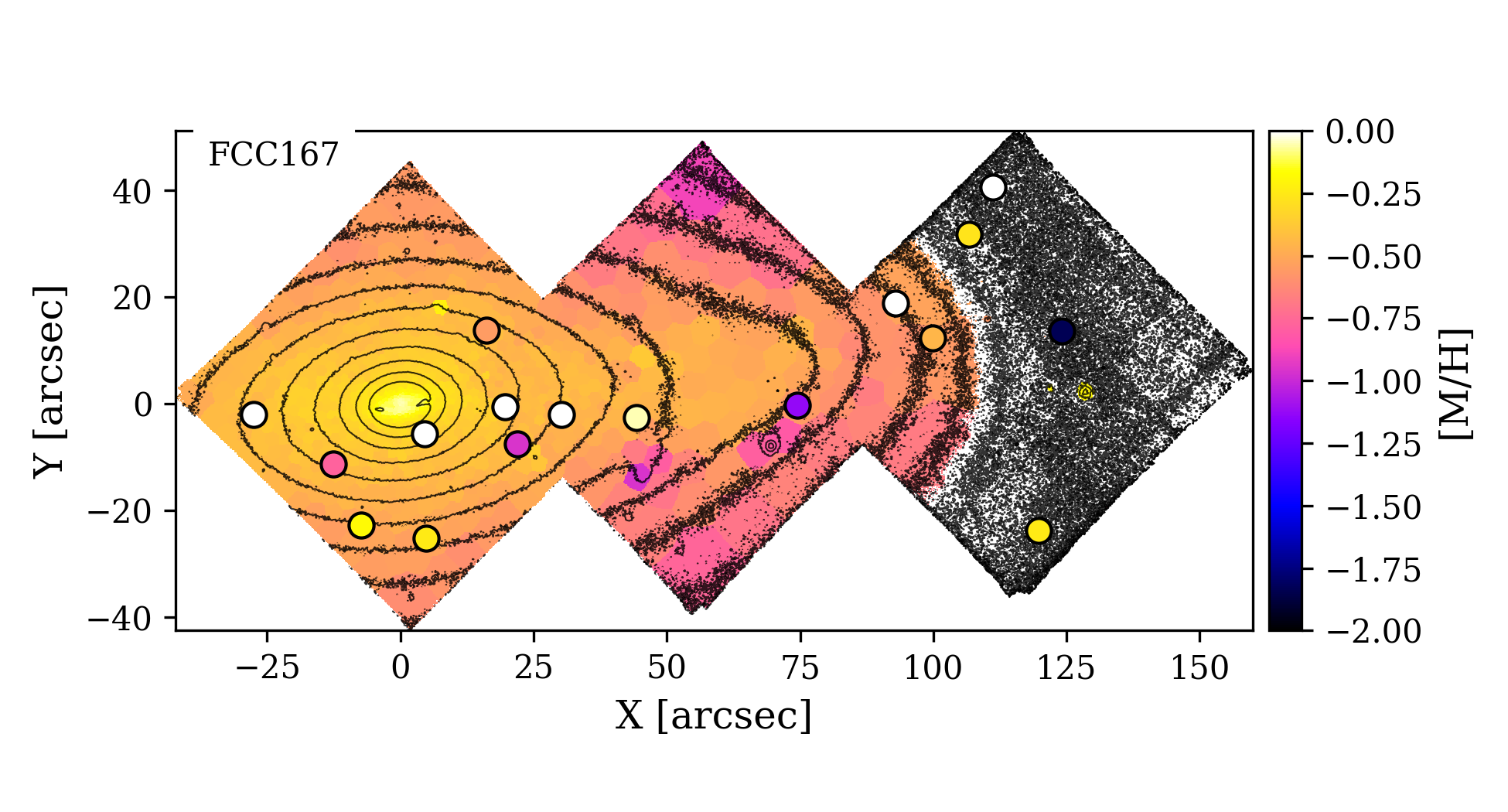}\\
\includegraphics[width=0.45\textwidth]{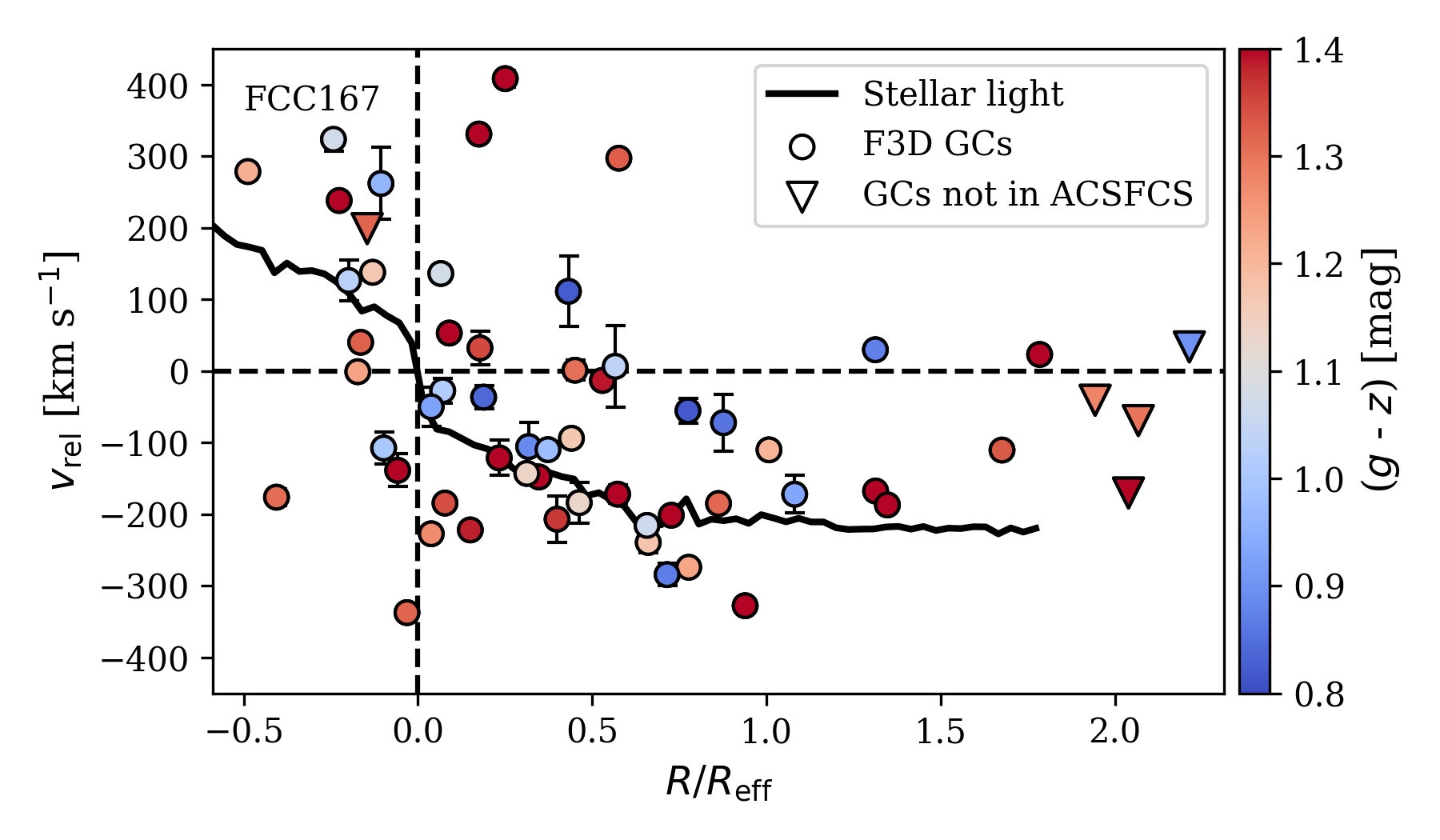}
\includegraphics[width=0.45\textwidth]{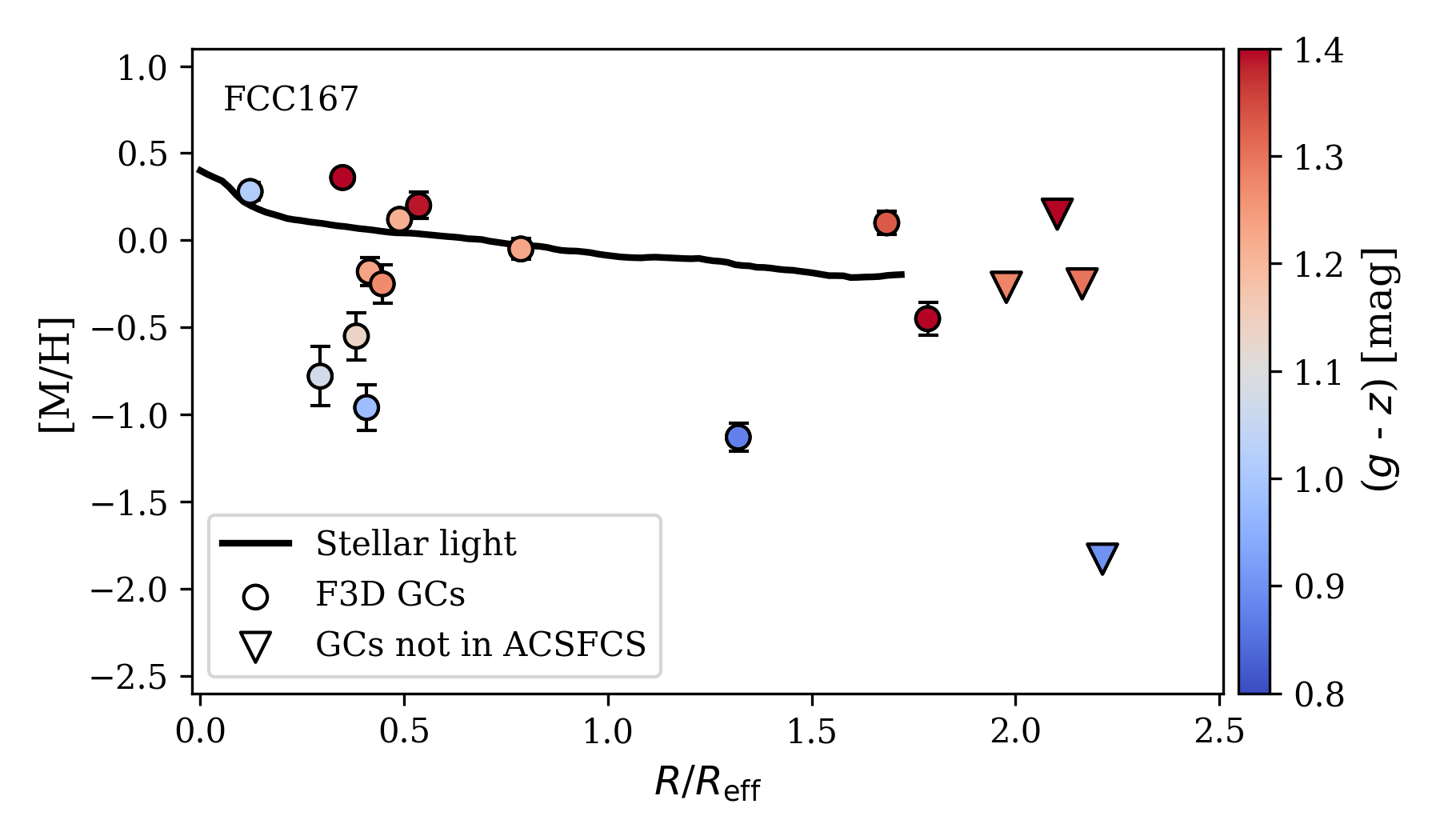}\\

\includegraphics[width=0.45\textwidth]{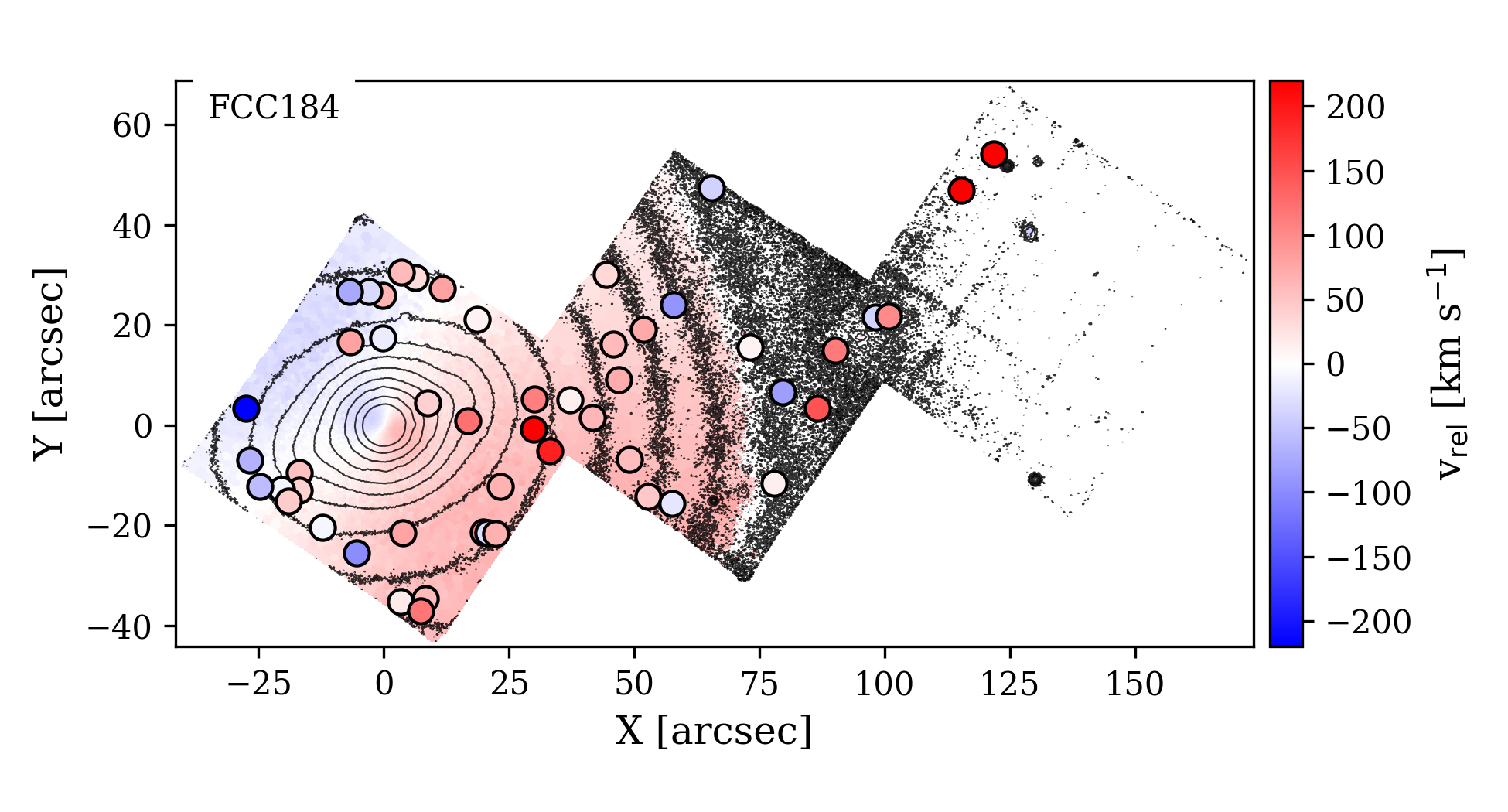}
\includegraphics[width=0.45\textwidth]{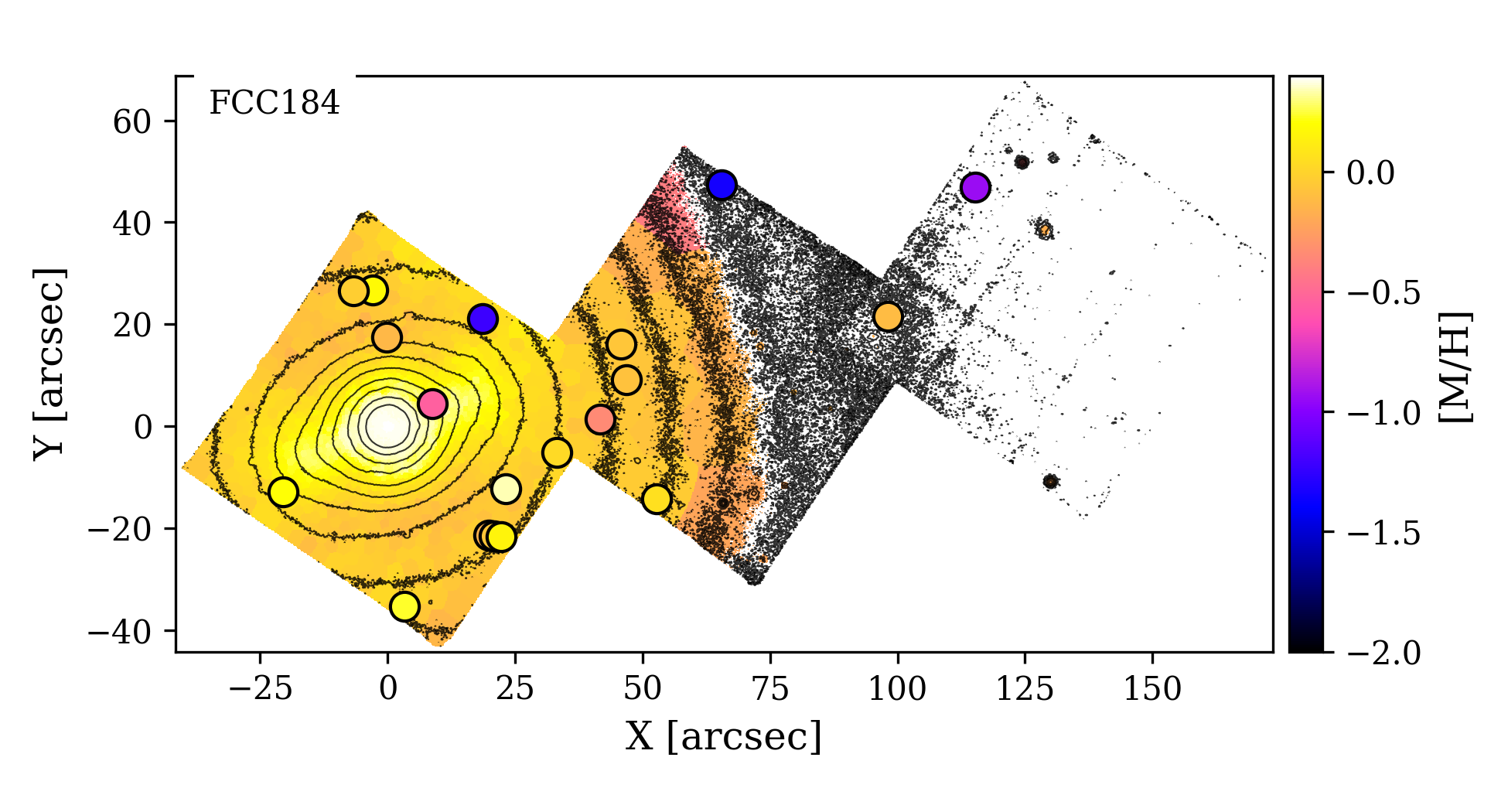}\\
\includegraphics[width=0.45\textwidth]{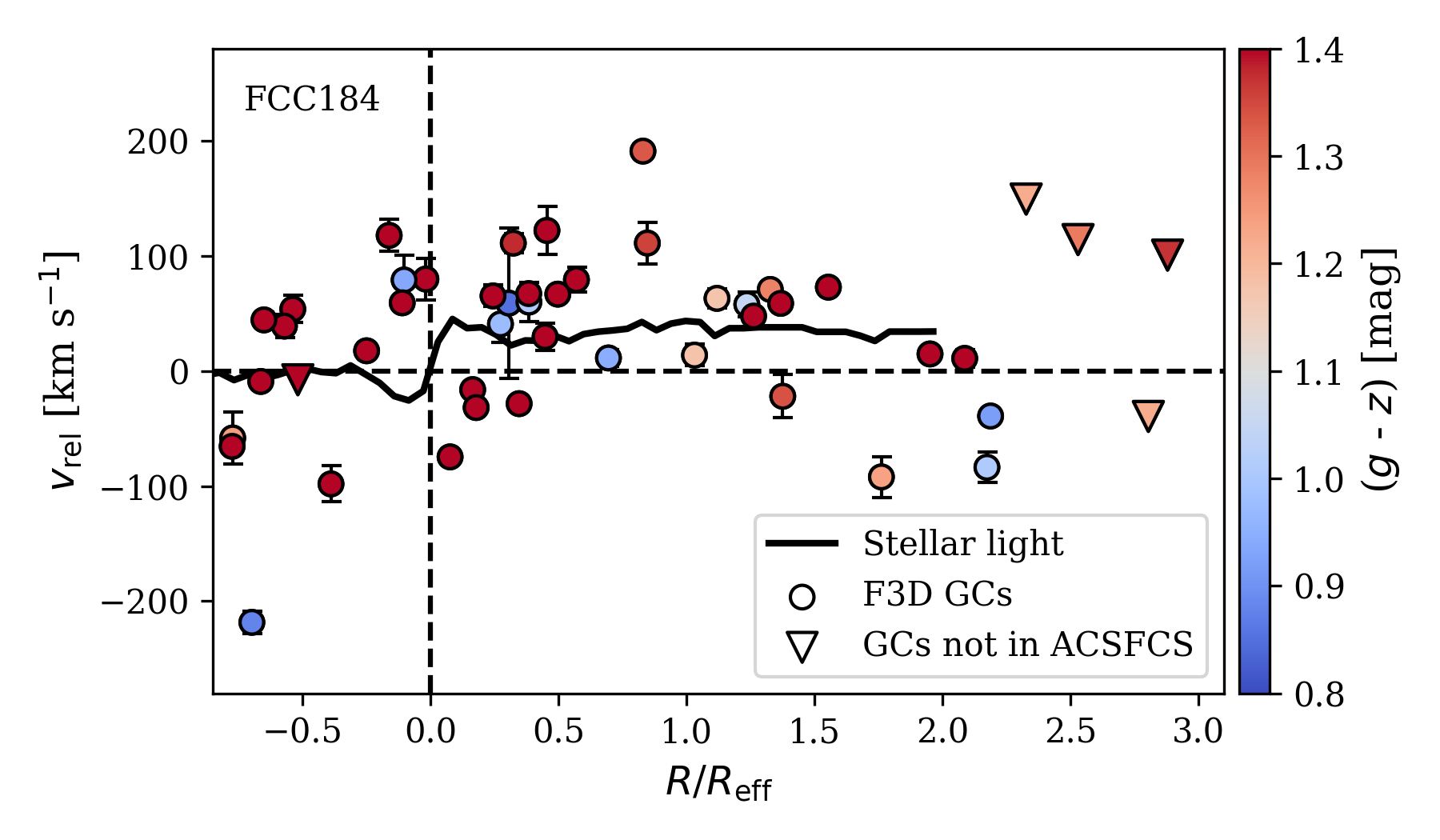}
\includegraphics[width=0.45\textwidth]{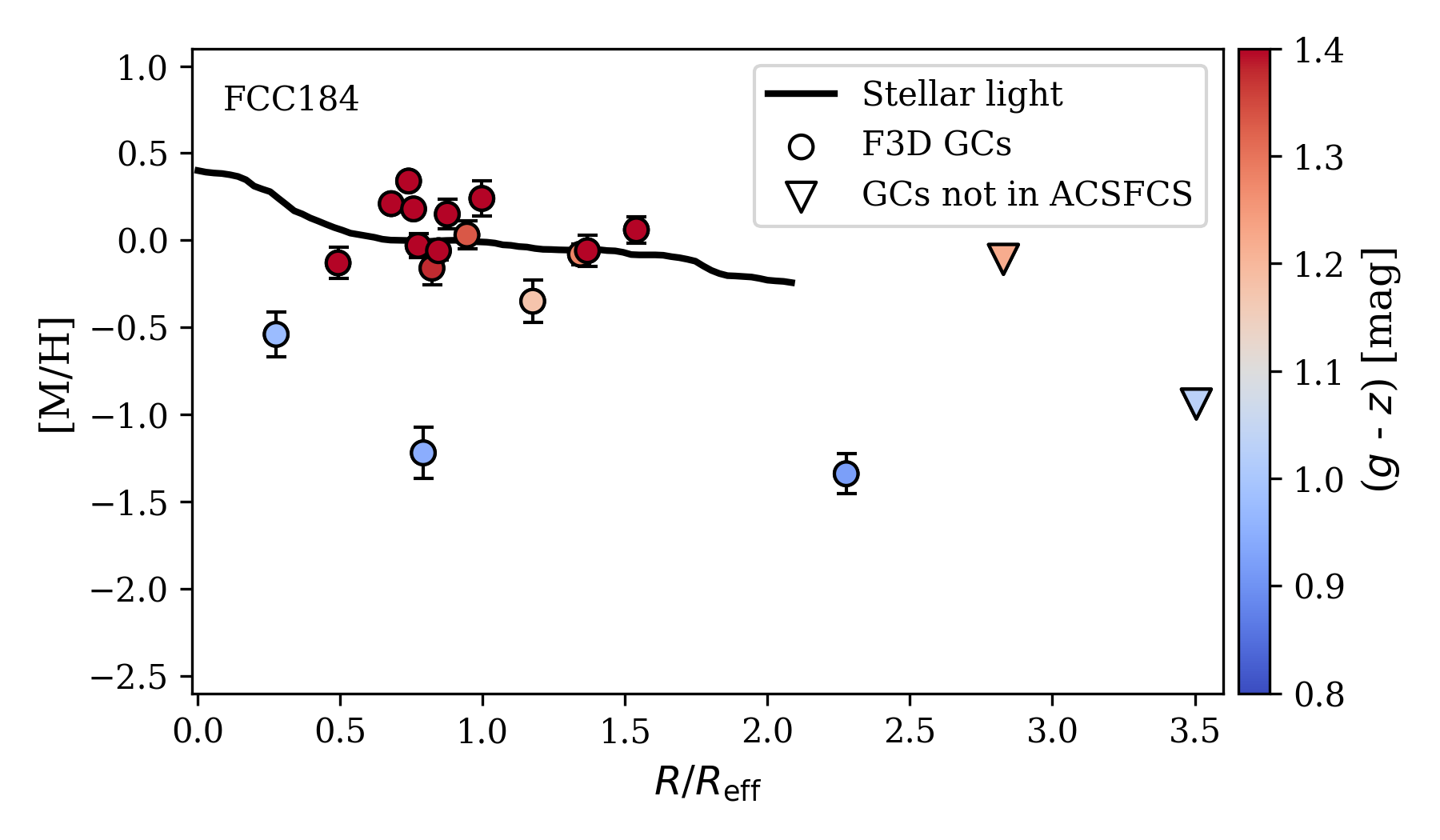}\\
\caption{Same as Fig. \ref{fig:maps1}, but for FCC\,167 and FCC\,184.}
\label{fig:maps6}
\end{figure*}

\end{document}